\newif\ifHEVEA \newif\ifDRAFT \newif\ifFINAL \newif\ifREVWS \newif\ifEXPLE \newif\ifACCEPTED

\newif\ifTWOCOLS

\TWOCOLSfalse

\HEVEAfalse
\DRAFTfalse
\REVWSfalse
\EXPLEfalse
\FINALtrue
\ACCEPTEDtrue

\TWOCOLStrue

\documentclass[sigconf]{acmart}

\newcommand{\abstractcount}[1]{}

\ifHEVEA
\else

\fi

\usepackage[normalem]{ulem} 

\usepackage{xcolor}

\ifREVWS
  \newcommand{\reviews}[1]{\textcolor{orange}{#1}} \else
  \newcommand{\reviews}[1]{} \fi

\ifEXPLE
  \newcommand{\example}[1]{\textcolor{yellow}{#1}} \else
  \newcommand{\example}[1]{} \fi

\newcommand{\white}[1]{\textcolor{white}{#1}} \newcommand{\black}[1]{\textcolor{black}{#1}} 

\ifDRAFT
\newcommand{\barre}[1]{\textcolor{lightgray}{\sout{#1}}} \newcommand{\barrePT}[1]{\textcolor{lightgray}{\sout{#1}}} 

  \newcommand{\remark}[1]{}

         \newcommand{\purple}[1]{\textcolor{purple}{#1}} 

\else
  \newcommand{\barre}[1]{} \newcommand{\barrePT}[1]{} 

  \ifFINAL
    \renewcommand{\reviews}[1]{} \newcommand{\remark}[1]{}    \else
    \newcommand{\remark}[1]{\textcolor{red}{\textsc{#1}}}    \fi

        \newcommand{\purple}[1]{#1} \fi

\ifDRAFT
  \usepackage{soul}
  \definecolor{bleuclair}{rgb}{0.7, 0.7, 1.0}
  \definecolor{rosepale}{rgb}{1.0, 0.7, 1.0}

\else

\fi

\def\ie{{i.e.,} }
\def\eg{{e.g.,} }

\newcommand{\ita}[1]{\textit{#1}}

\setcitestyle{notesep={: }}

\usepackage{eurosym}

\DeclareTextSymbol{\deg}{T1}{6}
\DeclareTextSymbol{\deg}{OT1}{23}

\newcommand\wordcount{
    \immediate\write18{texcount -sub=section \jobname.tex  | grep "Section" | sed -e 's/+.*//' | sed -n \thesection p > 'count.txt'}
(249
 words)}

\ifDRAFT
  \usepackage[yyyymmdd,12hr]{datetime} 
  \newcommand{\docversion}{<Document version: \today{} \currenttime{}>}
\else
  \newcommand\docversion[1]{}
\fi

\ifFINAL
  
\else
\fi

\usepackage{multirow}

\makeatletter\let\expandableinput\@@input\makeatother

\usepackage{rotating} 

\definecolor{coulD}{RGB}{52,167,255} 
\definecolor{coulT}{RGB}{51,255,146} 
\definecolor{coulO}{RGB}{255,190,51} 
\definecolor{coulC}{RGB}{255,51,178} 

\def\xd{\cellcolor{coulD}$\times$}
\def\xt{\cellcolor{coulT}$\times$}
\def\xo{\cellcolor{coulO}$\times$}
\def\xc{\cellcolor{coulC}$\times$}

\def\xa{\cellcolor{gray!95}$\times$} \def\xg{\cellcolor{gray!70}$\times$} \def\xi{\cellcolor{gray!50}$\times$}

\def\oo{\cellcolor{lightgray!50}-}

\DeclareUnicodeCharacter{202F}{FIX ME!!!!}

\usepackage{colortbl}

\usepackage{subcaption}

\usepackage[T1]{tipa}

\def\ie{{i.e.,} }
\def\eg{{e.g.,} }

\def\cen{\centering}
\def\raL{\raggedleft}

\DeclareTextSymbol{\deg}{T1}{6}
\DeclareTextSymbol{\deg}{OT1}{23}

\ifHEVEA
  \providecommand{\tabularnewline}{\\}
  \newcommand{\url}{}
  
  \newcommand{\g}[1]{#1}
  \newcommand{\q}[1]{#1}
\else
  \newcommand{\g}[1]{``#1''}
  \newcommand{\q}[1]{``#1''}
\fi

\usepackage{amsmath}

\def\myshorttitle{Defining a Role-Centered Terminology for Physical Representations and Controls}
\def\mytitle{Defining a Role-Centered Terminology for Physical Representations and Controls}

\def\mykeywords{terminology, linguistics, blended words, physical computing, physical user interfaces, graspable user interfaces, tangible user interfaces, embodied interfaces, taxonomy, classification}


\AtBeginDocument{}

\setcopyright{cc}
\copyrightyear{2025}
\acmYear{2025}

\acmConference[]{arXiv.org, Cornell University}{2025}{Ithaca, NY, USA}
\acmPrice{}
\acmISBN{}

\begin{document}

\title[\myshorttitle]{\mytitle}

\author{Guillaume Rivière}
\email{g.riviere@estia.fr}
\orcid{0000-0001-8390-9751}
\affiliation{\institution{Univ. Bordeaux, ESTIA Institute of Technology}
  \city{Bidart}
  \postcode{F-64210}
  \country{France}
}

\renewcommand{\shortauthors}{Riviere}

\begin{abstract}

Previous classifications advanced research through a better understanding of the field and the variety of tangible user interfaces and related physical user interfaces, especially by discretizing a degree of tangibility based on the specimens produced by the community over the years, since the conceptualization of Tangible User Interface initiated a research effort to deepen the exploration of the concept.
However, no taxonomy enables the classification of tangible user interfaces at the application level.
This article proposes to refine the description of tangible user interfaces' interactional components through a terminological approach.
The resulting terms are blended words, built from known words, that self-contain \textit{what} digital role is represented or controlled and \textit{how} it becomes physical.
This holistic terminology then enables the definition of applications' hallmarks and four classes of tangibility for applications, which surpass the description of physical user interface specimens' morphology by abstracting and discriminating specimens at the applicative level. 
The descriptiveness and holisticness of the new terminology, as well as the clustering and discriminative power of the limited number of four classes, are showed on a corpus of applicative tangible user interfaces' specimens from the literature.
Promising future work will benefit from the holistic terminology, the applications' hallmarks, and the tangibility classes, to describe applicative tangible user interfaces and related physical user interfaces to better understand the dozens of specimens that were produced by the field over three decades. Indeed, describing and classifying this whole set would deepen our understanding to provide tools for future developers and designers.

\end{abstract}


\maketitle

\docversion{}

\abstractcount{150}

\newcommand\WT[1]{{\footnotesize \white{#1}}}
\newcommand\BT[1]{{\footnotesize \black{#1}}}
\newcommand\Wt[1]{{\scriptsize \white{#1}}}
\newcommand\Bt[1]{{\scriptsize \black{#1}}}

\newcommand\sbullet[1][.5]{\mathbin{\vcenter{\hbox{\scalebox{#1}{$\bullet$}}}}} 

\def\numslotmachine{\#1}
\def\numcaad{\#2}
\def\numselfbuilder{\#3}
\def\nummarble{\#4}
\def\numpassiveprops{\#5}
\def\numgraspdraw{\#6}
\def\nummetadesk{\#7}
\def\numbuildit{\#8}
\def\numpinwheels{\#9}
\def\numvoodoodolls{\#10}
\def\nummediablocks{\#11}
\def\nummusicbottles{\#12}
\def\numurp{\#13}
\def\numsenseboard{\#14}
\def\numillclay{\#15}
\def\numaudiopad{\#16}
\def\numreactable{\#17}
\def\numipworkbench{\#18}
\def\numqueryshapes{\#19}
\def\numtuister{\#20}
\def\numiobrush{\#21}
\def\numpico{\#22}
\def\numarcheotui{\#23}
\def\numgeotui{\#24}
\def\numslurp{\#25}
\def\numrelief{\#26}
\def\numteegi{\#27}
\def\numsoundform{\#28}
\def\numcairnform{\#30}
\def\numrespire{\#29}
\def\numaxes{\#31}
\def\numcoda{\#32}
\def\numsablier{\#33}
 
\section{Introduction}

Conceptualization often precedes a community's focus on the same research path. Whereas precursor specimens---of what was later conceptualized as Tangible User Interfaces (TUIs)---arose in the mid-1970s and early 1980s \citetext{\citealp[Ch.~2]{shaer2010past}, \citealp{ullmer2022weaving}}, physical computing came to the forefront of Human--Computer Interaction (HCI) researchers' attention only fifteen years later \cite{wellner1991digitaldesk,hinckley1994passiveprops,fitzmaurice1995bricks}, after the concept of ubiquitous computing was conceptualized in the late 1980s \cite{weiser1991century,weiser1999century}.
Several works initiated physical computing's exploration, from ``passive real-world props'' \cite{hinckley1994passiveprops}, to ``graspable user interfaces'' \cite{fitzmaurice1995bricks}, to ``manipulative user interfaces'' \cite{harrison1998manipulative}, to ``embodied user interfaces'' \cite{fishkin1999embodied}, which were the growing roots of the same tree \cite{fishkin2004taxonomy}.
Likewise, the deeper exploration of the design and technology of tangible user interfaces began after they were conceptualized in 1997 \cite{ullmer1997metadesk}.
Since then, while the number of specimens has grown yearly, taxonomies, paradigms, and frameworks have emerged \citetext{\citealp{hoven2004tangible}, \citealp{mazalek2009framing}, \citealp[Ch.~5]{shaer2010past}}, defining terms to describe the roles of physical artifacts (\eg{} phicon \cite{ullmer1997metadesk}, token \cite{holmquist1999token,shaer2004tac,ullmer2005token}, and tangible representation \cite{ullmer2000emerging}).
However, no global terminology is shared in the literature, even if refining and conceptualizing tangible user interfaces' interactive components better could help refine the design space, deepen specimens' exploration, and broaden applications' possibilities---indeed, sharing a common vocabulary and the same concepts also helps aggregating research efforts on a common path.

Previous work provided isolated terms, common terms, and general terms, but none can distinguish all the entities involved in tangible user interfaces in a single namespace.
On the one hand, isolated terms are limited in the scopes and paradigms that define them (\eg{} Token And Constraints \cite{shaer2004tac}), and, therefore, remain separated in diverse namespaces.
On the other hand, some terms are found across several research works, but with inconsistent meanings (\eg{} \g{tokens} \cite{holmquist1999token,shaer2004tac,ullmer2005token}).
As a consequence, many researches are built upon general terms (\eg{} \g{tangibles,} \g{physical object,} and \g{tangible object} \cite{shaer2010past}), thus escaping terminological ambiguity and describing all physical artifacts as a whole set, without distinction.
However, such general terms remain broad and undescriptive. For instance, the term \g{tangibles} is plural and cannot name merely a single artifact.
The terms \g{physical object} and \g{tangible object} lend a new meaning to the term \g{object}, which is already overused in computer science \cite{shaer2004tac}. Finally, \g{objects} are already \g{physical} and \g{tangible} in the usual sense.

Furthermore, scopes other than pure tangible user interfaces resort to physical entities as representations of controls, but not especially for data representation (\eg{} in Augmented Reality \cite{cordeil2020axes} and Virtual Reality \cite{mahieux2022sablier} scopes).
The lack of precise terms can lead to consider any user interface resorting to physical objects as a \g{tangible user interface,} even if these objects are only a subpart of a user interface that does not represent data with physical bodiement, thus leading to confusion and misunderstanding (\eg{} before general frameworks, recognizing tangible user interfaces was mainly relying on ``\textit{I know one when I see one}'' approaches \cite{fishkin2004taxonomy}). Fishkin's taxonomy overcame these considerations by ``\textit{treating tangibility as a spectrum rather than a binary quantity}'' (\ie{} ``tangible'' and ``not tangible'' user interfaces) and organizing ``proof of concept'' specimens over a twenty-item discretized space \cite{fishkin2004taxonomy}. However, this taxonomy, which has contributed to better delineating and understanding the field, does not address tangible user interfaces at the applicative level (\ie{} user interfaces that gather and assemble components to meet the needs of final users).
Indeed, this taxonomy is mainly focused on analyzing data representations (\ie{} distance between data output and the inputs enabling data manipulation and edition, and the user actions' degree of analogy to perceive and manipulate data) and describes user interfaces as whole things (\ie{} without refining the components that compose applicative user interfaces). Except that such a global description requires determining what is ``the'' datum of the user interface, but this often becomes tricky in applicative user interfaces, which often involve multiple tasks, outputs, or users.
Although describing the morphology and meaning of tangible user interfaces globally enabled refining the concept of ``tangibility'' in physical user interfaces, developing applications of such user interfaces may require describing components separately and beyond data representations.

For instance, the Urp specimen (Urban Planning Workbench) \cite{benjoseph2001urp,ishii2002urp,underkoffler1999urp} melts, on the surface of a table, models of buildings with editing and navigation tools and with measurement objects, and all of these are augmented with video-projection to display information from the computational simulation. Describing this ``workbench'' application globally, as a whole thing, ignores all the subtleties of its functioning and, for example, the role played by intangible representations. As well, determining the distance between input and output fails when an applicative tangible user interface only conveys data (\ie{} no input).

In addition, terms describing intangible representations are underrepresented: all the previous terms---isolated, common, and general ones---focused nearly exclusively on detailing tangible representations (whereas many terms were defined for GUIs, such as icons, windows, or sliders).
Consequently, could we discriminate, in a single namespace, all the tangible, graspable, and intangible entities found in tangible user interfaces (which represent data physically) and in any other related physical user interfaces? Since previous work provided a profusion of specimens, could we use the refined terminology to better discriminate tangible user interfaces according to the description of their components? Consequently, could such a refined terminology of physical entities lead to a classification of applicative tangible user interfaces?

This article proposes a role-centered terminology that names physical entities by defining twelve terms that self-contain \textit{what} is represented or controlled and \textit{how}. This terminology serves two objectives: the first is to provide terms that better align with foundational definitions, thereby improving conceptual understanding; the second is to refine the description of the representations and controls involved in tangible and related physical user interfaces. To show the descriptive power of this terminology, this article successfully discriminates all the physical representations and controls found in a collection of applicative tangible user interfaces. Furthermore, this holistic discrimination also enables the classification of applicative tangible user interfaces into four tangibility classes. We envision and expect that such terminological and classification advances will serve as a foundation for future work to investigate the composition of previous applicative tangible user interfaces and related physical user interfaces in greater depth and to provide tools and toolkits for future designers.

After reviewing related work, this article introduces a new terminology based on a taxonomy with two axes, and then derives four tangibility classes of applicative tangible user interfaces from this terminology. Subsequently, the descriptiveness of the terminology is validated on a corpus of applicative tangible user interfaces, which are then clustered into the four classes. Then, previous terms from the literature are mapped to the new terminology, and the four classes are linked to previous classifications. Finally, before concluding, some limitations are highlighted, and directions are provided for future work.

\section{Related Work}\label{sec-related-work}

This section reviews the related work on terms for physical representations and controls in tangible user interfaces, the creation of new words in the usual language through the word blending, and taxonomies and classifications of tangible user interfaces.

\subsection{Terms for Physical Representations}\label{sec-related-terms}

The term \g{tangible} appeared early to describe a ``tangible model'' for 3D input in Computer-Aided Architectural Design (CAAD) \cite{aish1984caad} and ``tangible manipulation'' in the Digital Desk \cite{wellner1991digitaldesk} (but the latter was later renamed ``tactile interaction'' \cite{wellner1993digitaldesk}). Later, the terms ``graspable'' and ``tangible'' were introduced to define Graspable User Interfaces \cite{fitzmaurice1995bricks} and Tangible User Interfaces \cite{ullmer1997metadesk}, respectively. Finally, the term ``intangible'' appeared to name audio and video output modalities in the MCRit model\footnote{MCRit: ``Model-Control-Representation (intangible and tangible)'' \cite{ullmer2002phdthesis,ullmer2005token}, first known as MCRpd: ``Model-Control-Representation (physical and digital)'' \cite{ullmer2000emerging}.} \cite{ullmer2002phdthesis,ullmer2005token}. Whereas the term \g{tangible} is used close to dictionary definitions, the term \g{intangible} differs from the usual sense: it simply means ``not tangible.'' Therefore, intangible representations are those generated by the computer; they may disappear when the computer is turned off (\eg{} sound, video projection, and any audio and visual modalities or displays).

This section summarizes the usual terms used in the literature to name the interactional components of tangible user interfaces, categorized into general, isolated, and shared terms, as well as terms employed in specific toolkits and design tools.

\subsubsection{General Terms}\label{sec-terms-general}

Some terms are used in the literature to generically name the physical part of tangible user interfaces that embody digital information, including \q{physical artifacts} \cite{ullmer2000emerging}, \q{physical representations} \cite{ullmer2000emerging}, \q{tangible representations} \cite{ullmer2002phdthesis,ullmer2005token}, \q{tangibles} \cite{ullmer2002phdthesis}, and \q{tangible objects} \cite{shaer2010past}. Those terms are variations to name graspable and tangible interactional components broadly. However, the common use of the term ``tangible'' was initially defined for bodied data representations (\ie{} not for graspable anchors on tabletops). Finally, only one general term refers to neither graspable nor tangible components: \q{intangible representations} \cite{ullmer2002phdthesis,ullmer2005token}.

\subsubsection{Isolated Terms}\label{sec-terms-isolated}

The term \textit{passive real-world prop} \cite{hinckley1994passiveprops} is used to name bodied representations of tools and data that map sensed user actions directly to control digital data displayed on a distant screen. Props can be \textit{symbolic} \cite{ullmer2000emerging} (\eg{} a sphere \cite{hinckley1994passiveprops}) or \textit{iconic} \cite{ullmer2000emerging} (\eg{} a doll's head \cite{hinckley1994passiveprops}) data representations, combined with tools (\eg{} a slicing plate \cite{hinckley1994passiveprops}). The term \g{passive} refers to the system's ability to track and reproduce movements made on the props. We can understand the choice of this term with the first loop of tangible interaction \cite{ishii2008beyond}: props provide a first passive feedback loop before computation's feedback arises---the relative position of the props informs about cutting plane location without watching the screen (what is reinforced when using the doll's head \cite{hinckley1996thesis}).

The term \textit{physical handle} \cite{fitzmaurice1995bricks} (or \textit{phandle} \cite{ullmer1997metadesk}) refers to physical artifacts of general form (\eg{} bricks and pucks) that can be attached to intangible shapes displayed on a surface. Once the physical handle acquires an intangible shape, motions are transmitted to control the digital object (\eg{} position, orientation, and size).

The term \textit{phicon} \cite{ullmer1997metadesk} stands for \g{physical icons} of a specialized form that instantiates graphical user interface (GUI) components physically. Once put on a horizontal surface, phicons are linked to digital data (\eg{} a building \cite{ullmer1997metadesk}), and intangible data is updated to match phicons' physical states (\eg{} a displayed map is updated \cite{ullmer1997metadesk}).

\subsubsection{Shared Terms}\label{sec-terms-shared}

Some terms do not name the artifacts by what they are but by their role and appear across several research works. For example, the terms \g{token,} \g{container,} and \g{tool} \cite{holmquist1999token} define three roles in giving access to digital information. \textit{Tokens} are specialized artifacts whose physical properties match the represented digital information (\eg{} phicons \cite{ullmer1997metadesk}). \textit{Containers} are generic artifacts that can represent any type of information. Whereas tokens are continuously assigned to the same digital information, containers are associated with any digital information over time. Finally, \textit{tools} are physical artifacts that have computational functions, enabling them to act on the displayed data (\eg{} physical handles \cite{fitzmaurice1995bricks}). Beyond naming physical artifacts, access points to digital information are also named: digital information is rendered by \textit{information faucets} (\eg{} display devices, speakers, and tactile devices).
However, the \g{Token+Constraint} approach \cite{ullmer2005token} suggests that all those artifacts are ``tokens'': \g{containers} are ``symbolic tokens with permanent binding,'' \g{tokens} are ``iconic tokens with dynamic binding,'' and \g{tools} are ``tokens bound to operations'' \cite{ullmer2005token}.

The \g{Token+Constraint} approach \cite{ullmer2005token} reuses the term \g{token} in its usual sense, which is also its standard usage in computer science. The term \g{token} is then used in combination with the term \g{constraint,} where \textit{tokens} are physical objects representing digital information, and \textit{constraints} are physical regions bound to digital operations. Physical constraints structure the placement and combination of tokens, which systems can sense and interpret to trigger operations.

The term \textit{token} is also used in the usual sense in the \g{TAC Paradigm} \cite{shaer2004tac}, based on the aforementioned \g{Token+Constraint} approach \cite{ullmer2002phdthesis,ullmer2005token}. However, the TAC Paradigm first names physical objects composing a tangible user interface under the term \textit{pyfo} to move away from the various uses of the term \g{object.} Pyfo is a constructive and recursive term (not atomic): pyfos are made of pyfos (\eg{} the six sides of a box are also pyfos \cite{shaer2004tac}). As well, pyfos are not limited to whole graspable physical objects: pyfos are also intangible physical objects (\eg{} intangible constraints drawn on a surface, such as Senseboard's displayed grid, are pyfos \cite{jacob2002senseboard}). Thus, \textit{tokens} are graspable pyfos representing digital information or computational functions; \textit{constraints} are graspable or intangible pyfos that structure the behavior of tokens. Beyond physical representations, TAC Paradigm's terminology also includes terms for digital elements, such as the term \textit{variable}, which names digital information or computational functions. Whereas some variables can be bound to tokens, semantic variables cannot. Finally, systems behave according to a series of rules: association triplets between a token, a variable, and some constraints. These rules are referred to as \textit{TAC} (\ie{} Token And Constraints).

\subsubsection{Terms in Toolkits and Design Tools}

Toolkits and design tools also require terms to assign names to physical representations and controls, as well as to differentiate their roles.

The term \textit{phidgets} \cite{greenberg2001phidgets} stands for \g{physical widgets}: packaged, low-level devices and software architecture inspired by GUI toolkits to ease the building of physical user interfaces. The Phidgets toolkit provides physical primitives for controllers, sensors, and mechanical parts that are commercially available\footnote{Phidgets' website: \url{https://www.phidgets.com/}, last accessed 2022-06-09.}.

ASUR++ \cite{dubois2002asur++,dubois2003asur++} is a notation for describing the physical and digital entities of mobile mixed systems. This notation extends the ASUR notation for mixed reality systems and is usable for tangible user interfaces \cite{dubois2003asur++}. The notation defines two terms for physical objects, by distinction of two digital roles: \ita{tools} that support a task (R\textsubscript{tool}) and \ita{targeted objects} that are modified by a task (R\textsubscript{object}).

The ROSS API \cite{wu2012api} defines four nested levels to describe physical representations and controls. A strength of this low-level framework is that it goes beyond the description of physical objects (RObject) by denoting active spaces (RSpace) and surfaces (RSurface) that can sense interaction. However, even if input controls---such as buttons, switches, and sliders---are described (RControl), physical objects' roles---such as tools and data representations---are not describable by the API.

\subsubsection{Existent Terms' Limitations}\label{sec-limitations}

The previous terms are built from two main strategies: drawing a parallel with GUI elements (\eg{} ``icon,'' ``widget,'' ``handle,'' and ``container'') or naming their physical roles (\eg{} ``prop,'' ``token,'' and ``constraint''). At certain points, these approaches enable the description of low embodiment (\eg{} \g{phandles}) and higher embodiment (\eg{} \g{phicon}). However, these approaches fail to establish a global terminology that can describe all the components residing within tangible user interfaces. For example, \g{tokens} can equally name artifacts representing data or functions. Moreover, tangible user interfaces often comprise representations that are not ``purely tangible'' (\ie{} not representing data and not with high embodiment), but no terms distinguish a ``pure'' tangible user interface from other related physical user interfaces. Thereby, the lack of descriptive and discriminative power leads to the wide use of general terms to name any artifact indistinctly.

Some of the terms listed above come from everyday objects (\eg{} ``tokens,'' ``props,'' ``handles,'' ``faucets,'' ``bricks,'' or ``pucks''); some are blended words created specifically (\eg{} ``phandle,'' ``phicon,'' or ``phidget''); and some others associate two words (\eg{} ``physical artifact,'' ``tangible representation,'' or ``tangible object''). Most terms describe the artifacts but not how they relate to the digital world, which is suggested by a metaphor of their original use transposed to the digital world (\eg{} ``tokens,'' ``props,'' and ``faucets''). Only a few support the idea of digital entities coming to the physical world (such as ``phidgets'' and ``phicons'')---but they are isolated terms---and only a few express the coupling with a digital entity (\eg{} ``tangible representation'' and ``physical representation'').

As a general term, this article considers that ``physical representations'' comprise tangible, graspable, and intangible representations (as pyfos that also consider intangible representations as being physical)---the difference between ``tangible'' and ``graspable'' representations residing in the degree of embodiment: high and low, respectively. High embodiment occurs when some of the digital entities' representational characteristics are embedded in their representations (similarly to the \textit{specialized form} artifact definition \cite{fitzmaurice1997empirical}) and have iconic meaning \cite{ullmer2000emerging}. Low embodiment occurs when no digital entities' representational characteristics are embodied by their representations (similarly to the \textit{generic form} artifact's definition \cite{fitzmaurice1997empirical}) and that have symbolic meaning \cite{ullmer2000emerging}. This article also considers that the term ``artifacts'' refers to bodied objects, regardless of whether they are graspable or tangible. Finally, this article introduces a new holistic terminology that refines the descriptive level of the aforementioned physical representations by employing word-blending formation processes.

\subsection{Blended Words}\label{sec-related-blended-words}

Linguistics studies the evolution of everyday language over time, including word-formation processes (\eg{} derivations, abbreviations, and acronyms). Some new words are created by combining \textit{splinters} \cite{barrena2019splinters} excerpted from \textit{source words}. For example, \textit{blended words} concatenate a basis from a first word and a suffix from a second word, with phonemic overlap \cite{beliaeva2016blends,connolly2013innovation,gries2004shouldnt}. Linguists observed several possibilities and strategies of splinters excerption and combination that enter the definition of ``blended words'' that keep source words recognisable \cite{gries2004shouldnt,gries2006cognitive}. For instance, ``heliport'' (heli(copter) + (air)port), ``motel'' (mot(or) + (h)otel), and ``brunch'' (br(eakfast) + (l)unch) are blend words \cite{connolly2013innovation,gries2004shouldnt}. However, blends can follow some other word-formation processes. For example, \textit{clipping compounds} concatenate the bases of two words \cite{beliaeva2016blends,gries2004shouldnt,gries2006cognitive} (\eg{} ``fintech'' (fin(ancial) tech(nology)), ``sysadmin'' (sys(tem) + admin(istrator)), and ``hydrail'' (hyd(rogen) + rail(way))). Word blending enables creating new words of only 1, 2, or 3+ syllables \cite{connolly2013innovation}, thus creating some new vocabulary that designates concepts through single words, which also facilitates communication.

Blended words already exist in Computer Science and Human--Computer Interaction: for example, \g{pixel} (pic(ture)s + el(ement)) and \g{widget} (wi(ndow) + (ga)dget) are kinds of blended words that have become common in the usual language and are referenced in standard dictionaries. In the field of tangible user interfaces, several of the aforementioned terms are blended words, such as \g{phicon} (ph(ysical) + icon)  \cite{ullmer1997metadesk}, \g{phandle} (ph(ysical) + (h)andle) \cite{ullmer1997metadesk}, and \g{phidget} (ph(ysical) + (w)idget) \cite{greenberg2001phidgets}. 

This article employs word blend strategies to create new three-syllable terms that name physical entities found in tangible and related physical user interfaces.

\subsection{Taxonomies and Classifications}\label{sec-related-taxonomies}

The literature has resorted to taxonomies and classifications to characterize the conceptual space of physical user interfaces. Whereas the seminal work on graspable user interfaces proposed a conceptual space with thirteen axes \cite{fitzmaurice1995bricks}, Ishii later classified tangible user interfaces through eight genres \cite{ishii2008beyond}. About two dozen frameworks organized the conceptual space along different numbers of axes \citetext{\citealp{mazalek2009framing}, \citealp[Ch.~5]{shaer2010past}, \citealp[Ch.~3]{ullmer2022weaving}}.
For example, Fishkin's taxonomy of physicality (or tangibility) \cite{fishkin2004taxonomy} organized physical user interfaces as a twenty-item conceptual space along two axes: \textit{embodiment} (\ie{} the physical relationship between the input and the output of interaction loops) and \textit{metaphor} (\ie{} the digital meaning given to interaction loops with physical representations).
All those frameworks were organized by \textit{types}: abstracting, designing, and building; and by \textit{facets}: experiences, domains, physicality, interactions, and technologies \cite{mazalek2009framing}.

This article proposes a new taxonomy on two axes for the physical entities found in applicative tangible user interfaces and related physical user interfaces, based on the \textit{role} (\ie{} \textit{what}) and \textit{tangibility} (\ie{} \textit{how}) of the digital entities that are represented or controlled. A terminology is then built by taking the values of these two axes as source words for a blended-word formation process.

\section{Proposing a Terminology and a Classification}\label{sec-proposition}

This section introduces a new terminology that shifts the existing viewpoint from ``physical artifacts that are coupled to the digital world'' to a role-centered viewpoint of ``data coming to the physical world.'' Before defining the terminology, the definition of physical entities is refined, and then the concept of a tangible user interface is recalled. Finally, an abstraction of applicative tangible user interfaces and related physical user interfaces is proposed, along with their classification into four tangibility classes.

\subsection{Physical Interactional Entities}

Let us first define the set of physical interactional entities involved in tangible and related physical user interfaces, which the new terminology aims to name.
Indeed, the definitions in this article build upon the definition of ``Physical Interactional Entities,'' which are considered the highest level of abstraction of anything that is generated, animated, or sensed by computing resources and that is perceivable, editable, or manipulable by users. Such ``Interactional Entities'' are composable, couplable, and hybridable. ``Physical'' comprises matter (\eg{} objects, sides, and substances), sound (\eg{} audio, percussion, and noise), and any graphical item (\eg{} shapes, texts, and images). For example, the sides of a cube can be considered as entities, as well as the whole cube, depending on what is sensed, mapped, or represented.
Some other entities are added to this initial set that are not generated, nor animated, nor sensed by computing resources, but that support, hold, or guide entities from the initial set during interaction (\eg{} surfaces, slots, and trails).

\subsection{Back to Tangible User Interfaces' Concept}\label{sec-back-to-concept}

According to Ullmer and Ishii: \textit{``Tangible interfaces give physical form to digital information, employing physical artifacts both as representations and controls for computational media''} \cite{ullmer2000emerging}. This definition's phrasing reveals Ullmer and Ishii's datum-centered viewpoint on tangible user interfaces \cite[p. 19]{shaer2010past}. The adjective \g{tangible} should then be used to qualify data that becomes tangible, not to qualify artifacts or bodied objects (the latter is already ``tangible'' in the usual sense). This article claims we should then rather talk about ``tangible data,'' and no more about ``tangible artifacts'' or ``tangible objects.'' Thus, \textit{tangible data} are data that become tangible through artifacts (that give physical form to data): only those ``pure'' artifacts must be qualified as \textit{tangible representations} to match Ullmer and Ishii's definition strictly.
By extending this line of reasoning, this article proposes naming all artifacts according to their digital role---as instantiations of digital information---by providing meaning about how they represent digital entities.
This way, the following section defines a what--how terminology to name physical representations and controls from a role-centered viewpoint.

\subsection{A Role-Centered Terminology}\label{sec-proposing}

The new terminology uses blended words \cite{beliaeva2016blends,connolly2013innovation} formed from a taxonomy on ``what'' digital entities come into the digital world and ``how''.
The first part concerns \textit{what} digital entities are represented or controlled by physical representations: 

\begin{enumerate}
\item \textit{Data} refers to datum models and data states stored in computer memory and computed by the processor.
\item \textit{Tools} are instruments that can edit the data represented in the physical world (regardless of whether data representation is tangible, graspable, or intangible).
\item \textit{Operations} involve triggering an event or action in the computing system (\eg{} core operations \cite{ullmer2008core} include loading or saving data, loading an app, and validation).
\item Finally, \textit{constraints} structure the behavior of data representations, thus structuring mapped digital information. They can give meaning to artifacts by triggering actions or operations (\eg{} detecting a block's presence to trigger an action \cite{ullmer1999mediablocks,ullmer1998mediablocks}). Broadly, \g{constraints} affect physical objects' positioning (\eg{} table surface) before being sensed by the computer. \g{Constraints} are not always computer-generated: for example, some circles where to place artifacts can be video-projected onto a surface or drawn with paint.
\end{enumerate}

The second part concerns \textit{how} digital entities come to the physical world: 

\begin{enumerate}
\item Digital entities made \g{tangible} are represented by iconic artifacts of specialized form---with a high embodiment (of either or both representation and control).
\item Those that are made \g{graspable} are represented by symbolic artifacts of a generic form, with a low level of embodiment. Saying tangible or graspable is not saying better or worse, but simply different gradations of embodiment \cite{fishkin2004taxonomy} that correspond to different design and technology answers to different needs.
\item As well, digital entities made \g{intangible} are represented by volatile computer-generated physical phenomena that cannot be grasped and disappear once the computer is turned off (\eg{} display, video projection, or audio).
\end{enumerate}

From the seven source words mentioned above, the seven following splinters \cite{barrena2019splinters} are excerpted: four bases from \textit{what} words: \mbox{\g{dat-,}} \mbox{\g{tol-,}} \mbox{\g{op-,}} and \mbox{\g{const-}}; and three suffixes from \textit{how} words: \mbox{\g{-ible,}} \mbox{\g{-able,}} and \mbox{\g{-nible.}} Several attempts were required to select those splinters by a compromise between pronunciation, syllable count, and recognition of original words in the final terms (at least for experts aware of the terminology's construction). Blending those bases and suffixes thus provides twelve terms embedding meaning about \textit{what} and \textit{how} digital entities come to the physical world (see \autoref{tab::terms}).

The first term, \g{datible}---``datum is tangible''---better matches the meaning of the definition of ``purely tangible'' user interfaces, in contrast to previous terminologies. Moreover, none of the \purple{twelve} terms confuses words used for everyday objects. Finally, they provide a subtle and global descriptive power by naming four roles with three degrees in a unique namespace.
The following section suggests relying on the holisticness of the terminology to characterize applicative tangible user interfaces and related physical user interfaces through vertices.

\begin{table}
{\centering\small
\begin{tabular}{@{\hspace{1pt}}p{1.3cm}@{\hspace{2pt}}p{0.0cm}@{\hspace{6pt}}p{2.02cm}@{\hspace{6pt}}p{2.05cm}@{\hspace{9pt}}p{2.10cm}}
\cmidrule{3-5}
    & & \multicolumn{3}{l}{\textbf{\textit{How}}}           \tabularnewline
\cmidrule{1-1}    \cmidrule{3-5}
\textbf{\textit{What}} & & \it Tangible & \it Graspable & \it Intangible \tabularnewline
\cmidrule{1-1}    \cmidrule{3-5}

\it Datum &
& DATIBLE\vspace{-3pt}\linebreak{\scriptsize\textipa{["deItIb9l]}}\vspace{-3pt}\linebreak{\scriptsize `Datum is tangible'}& DATABLE\vspace{-3pt}\linebreak{\scriptsize\textipa{["deIt9b9l]}}\vspace{-3pt}\linebreak{\scriptsize `Datum is graspable'}& DATNIBLE\vspace{-3pt}\linebreak{\scriptsize\textipa{["deIt.nIb9l]}}\vspace{-3pt}\linebreak{\scriptsize `Datum is intangible'}\vspace{2pt}\tabularnewline

\it Tool &
& TOLIBLE\vspace{-3pt}\linebreak{\scriptsize\textipa{[tu:lIb9l]}}\vspace{-3pt}\linebreak{\scriptsize `Tool is tangible'}& TOLABLE\vspace{-3pt}\linebreak{\scriptsize\textipa{[tu:l9b9l]}}\vspace{-3pt}\linebreak{\scriptsize `Tool is graspable'}& TOLNIBLE\vspace{-3pt}\linebreak{\scriptsize\textipa{[tu:l.nIb9l]}}\vspace{-3pt}\linebreak{\scriptsize `Tool is intangible'}\vspace{2pt}\tabularnewline

\it Operation &
& OPIBLE\vspace{-3pt}\linebreak{\scriptsize\textipa{[OpIb9l]}}\vspace{-3pt}\linebreak{\scriptsize `Operation is tangible'}& OPABLE\vspace{-3pt}\linebreak{\scriptsize\textipa{[Op9b9l]}}\vspace{-3pt}\linebreak{\scriptsize `Operation is graspable'}& OPNIBLE\vspace{-3pt}\linebreak{\scriptsize\textipa{[Op.nIb9l]}}\vspace{-3pt}\linebreak{\scriptsize `Operation is intangible'}\vspace{2pt}\tabularnewline

\it Constraint &
& CONSTIBLE\vspace{-3pt}\linebreak{\scriptsize\textipa{["kOnstIb9l]}}\vspace{-3pt}\linebreak{\scriptsize `Constraint is tangible'}& CONSTABLE\vspace{-3pt}\linebreak{\scriptsize\textipa{["kOnst9b9l]}}\vspace{-3pt}\linebreak{\scriptsize `Constraint is graspable'}& CONSTNIBLE\vspace{-3pt}\linebreak{\scriptsize\textipa{["kOnst.nIb9l]}}\vspace{-3pt}\linebreak{\scriptsize `Constraint is intangible'}\tabularnewline

\cmidrule{1-1}    \cmidrule{3-5}
\end{tabular}
}
\caption{The twelve terms refining physical representations.}~\label{tab::terms}
\vspace{-9pt}
\end{table}

\subsection{An Abstraction Through Hallmarks}

We propose that applications could be abstracted by counting occurrences of physical representations and controls according to the \purple{twelve} terms discriminated by the holistic terminology. Such a characterization would then discriminate physical user interfaces at the application level, independently of their morphological and technological descriptions (\eg{} ``\textit{input is distant from output}'' \cite{fishkin2004taxonomy}, ``\textit{this interface's genre is tabletop}'' \cite{ishii2008beyond}, or ``\textit{input is sensed in 3D space by 2$\times$6DOF props}''). Therefore, this characterization could discriminate, for example, between two tabletop applications. Furthermore, this characterization could enable the comparison of applications within the same space, regardless of whether they are tabletop or ambient interfaces, or whether they have input that is distant from or nearby the output.

To this end, we define the two following vertices:
\begin{itemize}
\item Let the \textit{hallmark} of an application be the vertex of entities' count for each term, where terms are ordered first by \g{what} (\ie{} Datum, Tangible, Operation, then Constraint), and second by \g{how} (\ie{} Tangible, Graspable, then Intangible). The resulting hallmark is the following vertex: \newline
  (\textit{\#datible, \#datable, \#datnible, \#tolible, \#tolable, \#tolnible, \#opible, \#opable, \#opnible, \#constible, \#constable, \#constnible}). 
  \item Let the \textit{binary hallmark} of an application be the vertex built from the hallmark where components only indicate entities' presence for each term (\ie{} components' values greater than 1 are cut to 1). 
\end{itemize}

Some distances could be calculated between these vertices. However, this article primarily proposes using hallmarks to cluster applicative tangible user interfaces into distinct classes.

\subsection{Four Role-Centered Tangibility Classes}

We propose to cluster applicative tangible user interfaces according to how roles are represented, with a particular focus on the role of data, whose representation is determinant to the definition of tangible user interfaces. The four resulting tangibility classes are the following:

\begin{itemize}

\item In \textit{Class~I} specimens, data are represented only by tangible or graspable entities (\ie{} presence of either datibles or datables, but absence of datnibles). 

\item In \textit{Class~II} specimens, data are represented by intangible entities, as well as tangible or graspable entities (\ie{} presence of datnibles with either datibles or datables). 

\item In \textit{Class~III} specimens, data are represented only by intangible entities, which are coupled with the use of tangible or graspable tools (\ie{} presence of datnibles combined with tolibles or tolables, but absence of datibles and datables).

\item In \textit{Class~IV} specimens, only operations are provided, through tangible or graspable entities (\ie{} presence of opibles or opables, and data and tools are unknown or absent). This class provides specimens that are not restricted to a specific application and can thus control operations shared across applications (\ie{} core operations \cite{ullmer2008core}). 

\end{itemize}

The application hallmarks' patterns that correspond to these tangibility classes are reported in \autoref{tab::classes-definition}.
The following section illustrates the descriptive power of the twelve terms through examples from the literature.

\begin{table}
{\centering\small

\begin{tabular}{
      @{\hspace{0pt}}p{8mm}
      p{7mm}
      @{\hspace{0pt}}p{1mm}
      @{\hspace{0pt}}p{3mm}@{\hspace{2pt}}p{3mm}@{\hspace{2pt}}p{3mm}@{\hspace{0pt}}p{2pt}
      @{\hspace{0pt}}p{3mm}@{\hspace{2pt}}p{3mm}@{\hspace{2pt}}p{3mm}@{\hspace{0pt}}p{2pt}
      @{\hspace{0pt}}p{3mm}@{\hspace{2pt}}p{3mm}@{\hspace{2pt}}p{3mm}@{\hspace{0pt}}p{2pt}
      @{\hspace{0pt}}p{3mm}@{\hspace{2pt}}p{3mm}@{\hspace{3pt}}p{2mm}
      @{\hspace{2pt}}p{1mm}
    }
\midrule
& \multicolumn{16}{l}{\textit{\textbf{Hallmark pattern}}}\tabularnewline
\cmidrule{2-19}
& \textit{What:} && \multicolumn{3}{l}{\textit{D}} && \multicolumn{3}{l}{\textit{T}} && \multicolumn{3}{l}{\textit{O}} && \multicolumn{3}{l}{\textit{C}} & \tabularnewline
\cmidrule{4-6}\cmidrule{8-10}\cmidrule{12-14}\cmidrule{16-18}
\centering\textit{\textbf{Class}} & \textit{How:} &&\cen\it T &\cen\it G &\cen\it I &&\cen\it T &\cen\it G &\cen\it I &&\cen\it T &\cen\it G &\cen\it I &&\cen\it T &\cen\it G &\cen\it I \tabularnewline
\midrule

\centering I   && ( &\raL +, &\raL 0, &\raL 0, &&\raL  $\ast$, &\raL $\ast$, &\raL $\ast$, &&\raL  $\ast$, &\raL $\ast$, &\raL $\ast$, &&\raL  $\ast$, &\raL $\ast$, &\raL $\ast$ & ) \tabularnewline
\centering     && ( &\raL 0, &\raL +, &\raL 0, &&\raL  $\ast$, &\raL $\ast$, &\raL $\ast$, &&\raL  $\ast$, &\raL $\ast$, &\raL $\ast$, &&\raL  $\ast$, &\raL $\ast$, &\raL $\ast$ & ) \tabularnewline

\centering II  && ( &\raL +, &\raL 0, &\raL +, &&\raL  $\ast$, &\raL $\ast$, &\raL $\ast$, &&\raL  $\ast$, &\raL $\ast$, &\raL $\ast$, &&\raL  $\ast$, &\raL $\ast$, &\raL $\ast$ & ) \tabularnewline
\centering     && ( &\raL 0, &\raL +, &\raL +, &&\raL  $\ast$, &\raL $\ast$, &\raL $\ast$, &&\raL  $\ast$, &\raL $\ast$, &\raL $\ast$, &&\raL  $\ast$, &\raL $\ast$, &\raL $\ast$ & ) \tabularnewline

\centering III && ( &\raL 0, &\raL 0, &\raL +, &&\raL  0, &\raL +, &\raL $\ast$, &&\raL  $\ast$, &\raL $\ast$, &\raL $\ast$, &&\raL  $\ast$, &\raL $\ast$, &\raL $\ast$ & ) \tabularnewline
\centering     && ( &\raL 0, &\raL 0, &\raL +, &&\raL  +, &\raL 0, &\raL $\ast$, &&\raL  $\ast$, &\raL $\ast$, &\raL $\ast$, &&\raL  $\ast$, &\raL $\ast$, &\raL $\ast$ & ) \tabularnewline

\centering IV  && ( &\raL 0, &\raL 0, &\raL 0, &&\raL  0, &\raL 0, &\raL 0, &&\raL  +, &\raL 0, &\raL $\ast$, &&\raL  $\ast$, &\raL $\ast$, &\raL $\ast$ & ) \tabularnewline

\centering     && ( &\raL 0, &\raL 0, &\raL 0, &&\raL  0, &\raL 0, &\raL 0, &&\raL  0, &\raL +, &\raL $\ast$, &&\raL  $\ast$, &\raL $\ast$, &\raL $\ast$ & ) \tabularnewline

\midrule
  \end{tabular}\\
  \scriptsize{\it
    Note. What: `D' = Data, `T' = Tool, `O' = Operation, `C' = Constraint.\vspace{-2pt}\linebreak
    How: `T' = Tangible, `G' = Graspable, `I' = Intangible.\vspace{-2pt}\linebreak
    `+' = some representations (one or more). `$\ast$' = none or some representations.
  }
}
\caption{Representations in the four tangibility classes.}~\label{tab::classes-definition}
\vspace{-9pt}
\end{table}

\section{Illustrating on a Corpus of Applications}\label{sec-illustrating}

This section illustrates the usage of the new terminology by describing and classifying a corpus of applications from the literature.

\subsection{Specimens' Selection}

\def\tuigenresfootnotetext{Seven genres of tangible user interface applications were described in 2008 \cite{ishii2008beyond}:
(1)~tangible telepresence,
(2)~tangibles with kinetic memory,
(3)~constructive assembly,
(4)~tokens and constraints,
(5)~interactive surfaces---tabletop TUI,
(6)~continuous plastic TUI,
(7)~augmented everyday objects, and
(8)~ambient media, to which actuated TUI and shape-changing TUI must now be added.}

Because researchers have produced hundreds of tangible user interface specimens over the past few decades \cite{fleck2018cladistics}, the present collection is not meant to be exhaustive, but a rather representative sample. The specimen selection process is neither as rigorous as that in systematic literature reviews nor intended to provide reliable statistics. Instead, the objective is to illustrate the terminology coverage and usefulness by describing a variety of emblematic tangible user interface specimens and related physical user interfaces. Because research interests vary over the years, restricting the analysis to a specific date range could have led to the exclusion of some tangible user interface kinds; therefore, the entire research period, from the premise to the more recent work, was taken into account. The selection thus gathered various genres\footnote{\tuigenresfootnotetext} of tangible user interfaces that were studied during this entire period. Finally, the selection was filtered to retain work that showed real applications. Work merely demonstrating a new technology or a new kind of tangible user interface was not retained, as its naming could vary across applications. Indeed, knowing the mapping to the digital world is necessary because being physical is not enough to be a tangible representation: the same representation can be tangible in some applications but not in others.

\begin{figure*}[h]
  \centering
  \def\svgwidth{\textwidth} {\scriptsize \begingroup \makeatletter \providecommand\color[2][]{\errmessage{(Inkscape) Color is used for the text in Inkscape, but the package 'color.sty' is not loaded}\renewcommand\color[2][]{}}\providecommand\transparent[1]{\errmessage{(Inkscape) Transparency is used (non-zero) for the text in Inkscape, but the package 'transparent.sty' is not loaded}\renewcommand\transparent[1]{}}\providecommand\rotatebox[2]{#2}\newcommand*\fsize{\dimexpr\f@size pt\relax}\newcommand*\lineheight[1]{\fontsize{\fsize}{#1\fsize}\selectfont}\ifx\svgwidth\undefined \setlength{\unitlength}{853.20204847bp}\ifx\svgscale\undefined \relax \else \setlength{\unitlength}{\unitlength * \real{\svgscale}}\fi \else \setlength{\unitlength}{\svgwidth}\fi \global\let\svgwidth\undefined \global\let\svgscale\undefined \makeatother \begin{picture}(1,0.29137741)\lineheight{1}\setlength\tabcolsep{0pt}\put(0,0){\includegraphics[width=\unitlength,page=1]{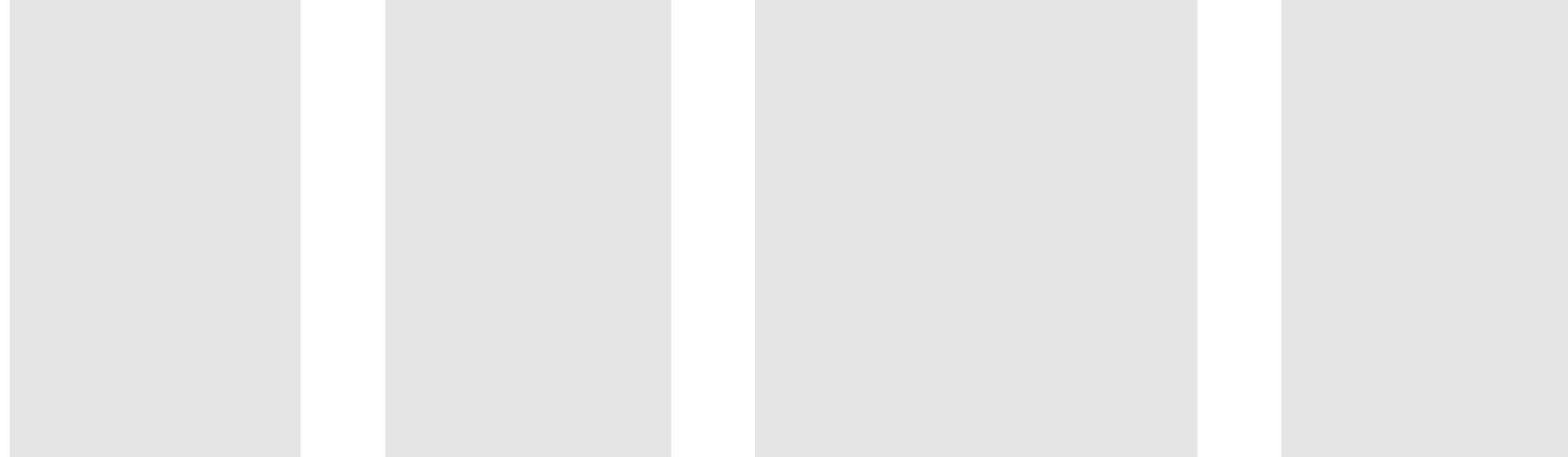}}\put(0.33463079,0.27736637){\color[rgb]{0,0,0}\makebox(0,0)[t]{\lineheight{1.25}\smash{\begin{tabular}[t]{c}\textbf{\textit{Browsing titles, abstracts,}}\\\textbf{\textit{and illustrations}}\end{tabular}}}}\put(0.09727722,0.27736637){\color[rgb]{0,0,0}\makebox(0,0)[t]{\lineheight{1.25}\smash{\begin{tabular}[t]{c}\textbf{\textit{Accessing sources}}\end{tabular}}}}\put(0.61488205,0.2775724){\color[rgb]{0,0,0}\makebox(0,0)[t]{\lineheight{1.25}\smash{\begin{tabular}[t]{c}\textbf{\textit{Extracting information from articles}}\end{tabular}}}}\put(0,0){\includegraphics[width=\unitlength,page=2]{selection-process.pdf}}\put(0.33645945,0.18987935){\color[rgb]{0,0,0}\makebox(0,0)[t]{\lineheight{1.25}\smash{\begin{tabular}[t]{c}Presenting a prototype?\end{tabular}}}}\put(0,0){\includegraphics[width=\unitlength,page=3]{selection-process.pdf}}\put(0.09900944,0.21421793){\color[rgb]{0,0,0}\makebox(0,0)[t]{\lineheight{1.25}\smash{\begin{tabular}[t]{c}State of the art \cite{shaer2010past}\end{tabular}}}}\put(0,0){\includegraphics[width=\unitlength,page=4]{selection-process.pdf}}\put(0.09900944,0.18093152){\color[rgb]{0,0,0}\makebox(0,0)[t]{\lineheight{1.25}\smash{\begin{tabular}[t]{c}List of specimens that are\\already known by the author\end{tabular}}}}\put(0,0){\includegraphics[width=\unitlength,page=5]{selection-process.pdf}}\put(0.09900943,0.13337536){\color[rgb]{0,0,0}\makebox(0,0)[t]{\lineheight{1.25}\smash{\begin{tabular}[t]{c}MIT TMG's webpage\end{tabular}}}}\put(0,0){\includegraphics[width=\unitlength,page=6]{selection-process.pdf}}\put(0.09900944,0.10046988){\color[rgb]{0,0,0}\makebox(0,0)[t]{\lineheight{1.25}\smash{\begin{tabular}[t]{c}ACM CHI 2013 to 2022:\\Sessions on "tangible"\end{tabular}}}}\put(0,0){\includegraphics[width=\unitlength,page=7]{selection-process.pdf}}\put(0.09900944,0.06003395){\color[rgb]{0,0,0}\makebox(0,0)[t]{\lineheight{1.25}\smash{\begin{tabular}[t]{c}ACM TEI 2013 to 2022:\\Proceedings\end{tabular}}}}\put(0,0){\includegraphics[width=\unitlength,page=8]{selection-process.pdf}}\put(0.33702779,0.13625777){\color[rgb]{0,0,0}\makebox(0,0)[t]{\lineheight{1.25}\smash{\begin{tabular}[t]{c}Applicative need?\end{tabular}}}}\put(0,0){\includegraphics[width=\unitlength,page=9]{selection-process.pdf}}\put(0.33702779,0.08996157){\color[rgb]{0,0,0}\makebox(0,0)[t]{\lineheight{1.25}\smash{\begin{tabular}[t]{c}Single user or\\co-presense conditions?\end{tabular}}}}\put(0,0){\includegraphics[width=\unitlength,page=10]{selection-process.pdf}}\put(0.62409164,0.09914081){\color[rgb]{0,0,0}\makebox(0,0)[t]{\lineheight{1.25}\smash{\begin{tabular}[t]{c}Genre and\\subgenre \\classification\end{tabular}}}}\put(0,0){\includegraphics[width=\unitlength,page=11]{selection-process.pdf}}\put(0.55552733,0.18108893){\color[rgb]{0,0,0}\makebox(0,0)[t]{\lineheight{1.25}\smash{\begin{tabular}[t]{c}Terminological\\description\end{tabular}}}}\put(0,0){\includegraphics[width=\unitlength,page=12]{selection-process.pdf}}\put(0.6915437,0.17974106){\color[rgb]{0,0,0}\makebox(0,0)[t]{\lineheight{1.25}\smash{\begin{tabular}[t]{c}Hallmark\\calculation\end{tabular}}}}\put(0,0){\includegraphics[width=\unitlength,page=13]{selection-process.pdf}}\put(0.90852647,0.27736638){\color[rgb]{0,0,0}\makebox(0,0)[t]{\lineheight{1.25}\smash{\begin{tabular}[t]{c}\textbf{\textit{Collecting specimens}}\end{tabular}}}}\put(0,0){\includegraphics[width=\unitlength,page=14]{selection-process.pdf}}\put(0.90856419,0.24794698){\color[rgb]{0,0,0}\makebox(0,0)[t]{\lineheight{1.25}\smash{\begin{tabular}[t]{c}First set\end{tabular}}}}\put(0,0){\includegraphics[width=\unitlength,page=15]{selection-process.pdf}}\put(0.90847377,0.13894575){\color[rgb]{0,0,0}\makebox(0,0)[t]{\lineheight{1.25}\smash{\begin{tabular}[t]{c}Second set\end{tabular}}}}\put(0,0){\includegraphics[width=\unitlength,page=16]{selection-process.pdf}}\put(0.90849721,0.20098546){\color[rgb]{0,0,0}\makebox(0,0)[t]{\lineheight{1.25}\smash{\begin{tabular}[t]{c}Does the first set cover\\the 12 terms?\end{tabular}}}}\put(0,0){\includegraphics[width=\unitlength,page=17]{selection-process.pdf}}\put(0.90849721,0.09161656){\color[rgb]{0,0,0}\makebox(0,0)[t]{\lineheight{1.25}\smash{\begin{tabular}[t]{c}Clustering by hallmarks\\and sorting by genres\end{tabular}}}}\put(0,0){\includegraphics[width=\unitlength,page=18]{selection-process.pdf}}\put(0.90847377,0.01988226){\color[rgb]{0,0,0}\makebox(0,0)[t]{\lineheight{1.25}\smash{\begin{tabular}[t]{c}Third set\end{tabular}}}}\put(0,0){\includegraphics[width=\unitlength,page=19]{selection-process.pdf}}\put(0.92429674,0.16531702){\color[rgb]{0,0,0}\makebox(0,0)[t]{\lineheight{1.25}\smash{\begin{tabular}[t]{c}Yes\end{tabular}}}}\put(0.78928418,0.20272597){\color[rgb]{0,0,0}\makebox(0,0)[t]{\lineheight{1.25}\smash{\begin{tabular}[t]{c}No\end{tabular}}}}\put(0,0){\includegraphics[width=\unitlength,page=20]{selection-process.pdf}}\put(0.91444564,0.05737786){\color[rgb]{0,0,0}\makebox(0,0)[lt]{\lineheight{1.14999998}\smash{\begin{tabular}[t]{l}Orthogonality?\\Heterogeneity?\end{tabular}}}}\put(0.45470276,0.15877385){\color[rgb]{0,0,0}\makebox(0,0)[t]{\lineheight{1.25}\smash{\begin{tabular}[t]{c}If yes,\\inclusion\end{tabular}}}}\end{picture}\endgroup  }
  \caption{The selection process of applicative specimens of tangible user interfaces and related physical user interfaces.}
  \Description{The diagram describes four successive steps: (1) Accessing sources, (2) Browsing titles, abstracts, and illustrations, (3) Extracting information from articles, and (4) Collecting specimens. The first step presents the five sources of articles. The second step sifts the articles to ensure they: (a) present a prototype, (b) answer to applicative needs, and (c) respond to single user or co-presense interaction conditions. The third step process the articles to provide applications' terminological description, hallmarks, and genre and subgenre classification. The fourth step is iterated to pump up articles until all criteria are reached: terminological coverage, hallmarks' orthogonality, and genres' heterogeneity.}
  \label{fig:selection-process}
\end{figure*}

The selection process is summarized in \autoref{fig:selection-process}. Several sources were browsed. First, a previous state of the art \cite{shaer2010past} and some specimens already known to the author (\eg{} some specimens that were experienced during demonstration sessions at conferences over time). Second, the projects page\footnote{MIT Tangible Media Group's projects page: \url{https://tangible.media.mit.edu/projects/}, last accessed 2022-08-03.} of the MIT Tangible Media Group, which is certainly the most prolific laboratory in this field (accounts for 45.5\% of the final collection's provenance). Third, to obtain even more recent work, sessions with `tangible' in their title from the last ten CHI conference proceedings (from 2013 to 2022 venues). Finally, the last ten TEI conference proceedings from 2013 to 2022 were also browsed. Specimens regarding remote computer-supported conditions were rejected.

Once many examples were added to the collection, the terms' coverage was computed to check whether the whole terminology is relevant. Next, more examples were selected and analyzed until the full coverage of the whole terminology's terms could be reached. \autoref{tab::terms-count} shows the number of physical representations gathered across the twelve terms.

\begin{table}
{\centering\small
  \begin{tabular}{p{7mm}p{7mm}p{1mm}@{\hspace{-2pt}}p{7mm}p{7mm}p{1mm}@{\hspace{-2pt}}p{7mm}p{7mm}p{1mm}@{\hspace{-2pt}}p{7mm}p{7mm}}
\cmidrule{1-2}\cmidrule{4-5}\cmidrule{7-8}\cmidrule{10-11}
\textit{Term} & \textit{Count} && \textit{Term} & \textit{Count} && \textit{Term} & \textit{Count} && \textit{Term} & \textit{Count}\tabularnewline
\cmidrule{1-2}\cmidrule{4-5}\cmidrule{7-8}\cmidrule{10-11}
Datible  &\raL 14 &&  Tolible  &\raL  6 &&  Opible  &\raL  8 &&  Constible  &\raL  3\tabularnewline 
Datable  &\raL 11 &&  Tolable  &\raL 17 &&  Opable  &\raL 12 &&  Constable  &\raL 19\tabularnewline
Datnible &\raL 38 &&  Tolnible &\raL  8 &&  Opnible &\raL  7 &&  Constnible &\raL  2\tabularnewline 
\cmidrule{1-2}\cmidrule{4-5}\cmidrule{7-8}\cmidrule{10-11}

\end{tabular}
}
\caption{Collection's physical representation count by terms.}~\label{tab::terms-count}
\vspace{-9pt}
\end{table}

\def\hmarkSlotMachine{(0, 1, 1, 0, 0, 0, 0, 2, 0, 0, 1, 0)}
\def\bhmarkSlotMachine{(0, 1, 1, 0, 0, 0, 0, 1, 0, 0, 1, 0)}
\def\hmarkCAADxDModellingSystem{(0, 1, 2, 0, 0, 0, 0, 0, 0, 0, 0, 0)}
\def\bhmarkCAADxDModellingSystem{(0, 1, 1, 0, 0, 0, 0, 0, 0, 0, 0, 0)}
\def\hmarkSelfBuilderModel{(1, 0, 2, 0, 0, 0, 0, 0, 0, 0, 1, 0)}
\def\bhmarkSelfBuilderModel{(1, 0, 1, 0, 0, 0, 0, 0, 0, 0, 1, 0)}
\def\hmarkMarbleAnsweringMachine{(1, 0, 1, 0, 0, 0, 0, 1, 0, 0, 1, 0)}
\def\bhmarkMarbleAnsweringMachine{(1, 0, 1, 0, 0, 0, 0, 1, 0, 0, 1, 0)}
\def\hmarkHeadProp{(1, 0, 1, 1, 0, 0, 0, 2, 0, 0, 0, 0)}
\def\bhmarkHeadProp{(1, 0, 1, 1, 0, 0, 0, 1, 0, 0, 0, 0)}
\def\hmarkGraspDraw{(0, 0, 1, 0, 1, 0, 1, 1, 0, 0, 1, 0)}
\def\bhmarkGraspDraw{(0, 0, 1, 0, 1, 0, 1, 1, 0, 0, 1, 0)}
\def\hmarkTangibleGeospace{(1, 0, 2, 0, 1, 0, 1, 0, 0, 0, 1, 0)}
\def\bhmarkTangibleGeospace{(1, 0, 1, 0, 1, 0, 1, 0, 0, 0, 1, 0)}
\def\hmarkBuildIT{(0, 0, 3, 0, 1, 0, 0, 0, 2, 0, 1, 0)}
\def\bhmarkBuildIT{(0, 0, 1, 0, 1, 0, 0, 0, 1, 0, 1, 0)}
\def\hmarkPinwheels{(N, 0, 0, 0, 0, 0, 0, 0, 0, 0, 0, 0)}
\def\bhmarkPinwheels{(N, 0, 0, 0, 0, 0, 0, 0, 0, 0, 0, 0)}
\def\hmarkVoodooDolls{(0, 1, 1, 0, 0, 0, 0, 0, 0, 0, 0, 0)}
\def\bhmarkVoodooDolls{(0, 1, 1, 0, 0, 0, 0, 0, 0, 0, 0, 0)}
\def\hmarkmediaBlocks{(0, 0, 0, 0, 0, 0, 0, 1, 0, 0, 1, 0)}
\def\bhmarkmediaBlocks{(0, 0, 0, 0, 0, 0, 0, 1, 0, 0, 1, 0)}
\def\hmarkmusicBottles{(0, 1, 1, 0, 0, 0, 1, 0, 0, 0, 1, 1)}
\def\bhmarkmusicBottles{(0, 1, 1, 0, 0, 0, 1, 0, 0, 0, 1, 1)}
\def\hmarkUrp{(2, 0, 2, 2, 2, 0, 0, 2, 2, 0, 1, 0)}
\def\bhmarkUrp{(1, 0, 1, 1, 1, 0, 0, 1, 1, 0, 1, 0)}
\def\hmarkSenseboard{(0, 1, 1, 0, 2, 0, 0, 0, 0, 0, 1, 1)}
\def\bhmarkSenseboard{(0, 1, 1, 0, 1, 0, 0, 0, 0, 0, 1, 1)}
\def\hmarkIlluminatingClay{(1, 0, 3, 0, 0, 1, 0, 0, 0, 0, 1, 0)}
\def\bhmarkIlluminatingClay{(1, 0, 1, 0, 0, 1, 0, 0, 0, 0, 1, 0)}
\def\hmarkAudioPad{(0, 1, 1, 0, 1, 0, 0, 0, 1, 0, 1, 0)}
\def\bhmarkAudioPad{(0, 1, 1, 0, 1, 0, 0, 0, 1, 0, 1, 0)}
\def\hmarkReacTable{(0, 1, 3, 0, 4, 0, 0, 0, 0, 0, 1, 0)}
\def\bhmarkReacTable{(0, 1, 1, 0, 1, 0, 0, 0, 0, 0, 1, 0)}
\def\hmarkIPNetworkDesignWorkbench{(0, 0, 2, 0, 2, 4, 0, 0, 2, 0, 1, 0)}
\def\bhmarkIPNetworkDesignWorkbench{(0, 0, 1, 0, 1, 1, 0, 0, 1, 0, 1, 0)}
\def\hmarkQueryShapes{(0, 1, 2, 0, 0, 0, 0, 0, 0, 0, 0, 0)}
\def\bhmarkQueryShapes{(0, 1, 1, 0, 0, 0, 0, 0, 0, 0, 0, 0)}
\def\hmarkTUISTER{(0, 0, 0, 0, 0, 0, 1, 0, 0, 0, 0, 0)}
\def\bhmarkTUISTER{(0, 0, 0, 0, 0, 0, 1, 0, 0, 0, 0, 0)}
\def\hmarkIOBrush{(0, 0, 1, 1, 0, 0, 1, 0, 0, 0, 0, 0)}
\def\bhmarkIOBrush{(0, 0, 1, 1, 0, 0, 1, 0, 0, 0, 0, 0)}
\def\hmarkPICO{(0, 1, 1, 0, 0, 0, 0, 0, 0, 3, 2, 0)}
\def\bhmarkPICO{(0, 1, 1, 0, 0, 0, 0, 0, 0, 1, 1, 0)}
\def\hmarkGeoTUI{(0, 0, 1, 1, 1, 1, 0, 1, 0, 0, 1, 0)}
\def\bhmarkGeoTUI{(0, 0, 1, 1, 1, 1, 0, 1, 0, 0, 1, 0)}
\def\hmarkSlurp{(0, 0, 0, 0, 0, 0, 1, 0, 0, 0, 0, 0)}
\def\bhmarkSlurp{(0, 0, 0, 0, 0, 0, 1, 0, 0, 0, 0, 0)}
\def\hmarkRelief{(1, 0, 0, 0, 0, 0, 0, 0, 0, 0, 0, 0)}
\def\bhmarkRelief{(1, 0, 0, 0, 0, 0, 0, 0, 0, 0, 0, 0)}
\def\hmarkTeegi{(2, 0, 1, 1, 1, 2, 0, 0, 0, 0, 1, 0)}
\def\bhmarkTeegi{(1, 0, 1, 1, 1, 1, 0, 0, 0, 0, 1, 0)}
\def\hmarkSoundFORMS{(1, 0, 1, 0, 0, 0, 0, 1, 0, 0, 0, 0)}
\def\bhmarkSoundFORMS{(1, 0, 1, 0, 0, 0, 0, 1, 0, 0, 0, 0)}
\def\hmarkCairnFORM{(0, 1, 0, 0, 0, 0, 0, 0, 0, 0, 0, 0)}
\def\bhmarkCairnFORM{(0, 1, 0, 0, 0, 0, 0, 0, 0, 0, 0, 0)}
\def\hmarkreSpire{(1, 0, 0, 0, 0, 0, 0, 0, 0, 0, 0, 0)}
\def\bhmarkreSpire{(1, 0, 0, 0, 0, 0, 0, 0, 0, 0, 0, 0)}
\def\hmarkEmbodiedAxes{(0, 0, 0, 0, 0, 0, 1, 0, 0, 0, 0, 0)}
\def\bhmarkEmbodiedAxes{(0, 0, 0, 0, 0, 0, 1, 0, 0, 0, 0, 0)}
\def\hmarkCoDa{(1, 0, 2, 0, 1, 0, 0, 0, 0, 0, 1, 0)}
\def\bhmarkCoDa{(1, 0, 1, 0, 1, 0, 0, 0, 0, 0, 1, 0)}
\def\hmarkSABLIER{(0, 0, 0, 0, 0, 0, 1, 0, 0, 0, 0, 0)}
\def\bhmarkSABLIER{(0, 0, 0, 0, 0, 0, 1, 0, 0, 0, 0, 0)}
 
\def\footnoteClusters{\footnote{
The three clusters' hallmarks are
and \hmarkRelief{}:
Relief (\numrelief{}),
and reSpire (\numrespire{});
\hmarkCAADxDModellingSystem{}:
CAAD 3D Modelling System (\numcaad{})
and Query Shapes (\numqueryshapes{});
\hmarkTUISTER{}:
TUISTER (\numtuister{}),
Slurp (\numslurp{}),
Embodied Axes (\numaxes{}),
and SABLIER (\numsablier{}).
}}

\def\footnoteClustersBinary{\footnote{
Pinwheels (\numpinwheels{}) joins the cluster of Relief (\numrelief{}) and reSpire (\numrespire{}) because sharing the same binary hallmark: \bhmarkRelief{}.
Voodoo Dolls (\numvoodoodolls{}) joins the cluster of CAAD 3D Modelling System (\numcaad{}) and Query Shapes (\numqueryshapes{}) because having the same binary hallmark: \bhmarkCAADxDModellingSystem{}.
}}

The orthogonality of the specimens across the terminology was also checked to assess the collection's diversity. To this end, the hallmarks and binary hallmarks of the applications were computed.

\subsection{Corpus Description}

Tables \ref{tab::illustration1}, \ref{tab::illustration2}, and \ref{tab::illustration3} show the resulting naming of the \purple{145} physical entities from the collection of applications.

The four roles enabled sorting all of these physical entities. Data is the most prevalent role, accounting for 43\% of the physical entities. However, 57\% of the physical entities are not data: 21\% are tools, 19\% are operations, and 17\% are constraints. Therefore, overemphasizing design or research on data representations would overlook a significant portion of the entities encountered in applicative tangible user interfaces.
Furthermore, the terminology enabled the naming of all of the physical entities found in the \purple{33} applications. Being able to distinguish and designate entities might be useful to designers. For example, \purple{six} of the applications are depicted in \autoref{fig:illustration} with their what--how naming.

\ifTWOCOLS
 \def\figwfactor{0.495}
 \def\fighspace{1pt}
 \def\figvspace{12pt}
\else
 \def\figwfactor{0.490}
 \def\fighspace{6pt}
 \def\figvspace{6pt}
\fi

\newcommand{\license}{
{\tiny
  Licenses:
  (b) Castle 3D model: Ramon Duran, \href{https://creativecommons.org/publicdomain/zero/1.0/}{CC0};
  (c) Doll's head 3D model: David Gilman, \href{https://creativecommons.org/licenses/by/4.0/}{CC BY 4.0} -- Brain MRI picture: Rudy Liggett, \href{https://creativecommons.org/publicdomain/zero/1.0/}{CC0};
  (d) Vase 3D model: Robert Tolliver, \href{https://creativecommons.org/licenses/by/4.0/}{CC BY 4.0};\linebreak
  (f) Topographic map: Gordon Dylan Johnson, \href{https://creativecommons.org/publicdomain/zero/1.0/}{CC0};
  (c,d,e,f) Chair 3D model: Zach Steindler, \href{https://opensource.org/licenses/MIT}{MIT License}.
  }
}

\begin{figure*}[ht!]
  \centering
\begin{subfigure}[b]{\figwfactor\linewidth}
    \centering
    \includegraphics[width=\linewidth,width=\linewidth]{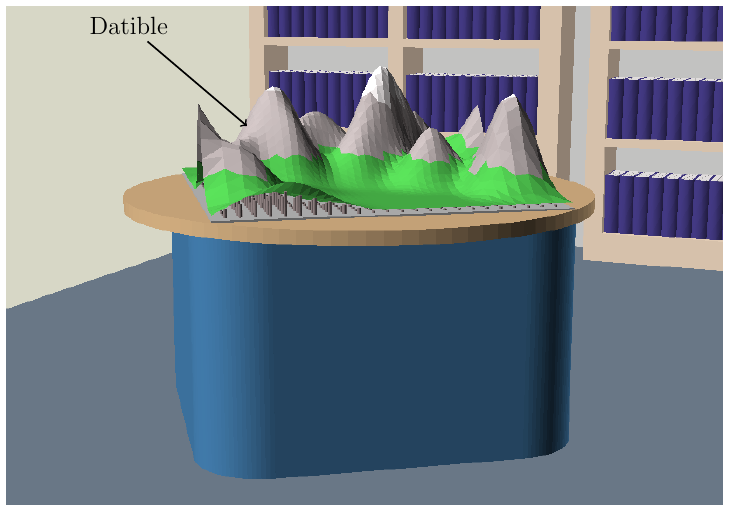}
    \caption{Relief: Terrain dynamic exploration (\numrelief{}).}
    \label{fig:illustration:relief}
    \Description{A table's dynamic surface (datible) displays mountain reliefs colored according to elevation.}
  \end{subfigure}\hspace{\fighspace}
  \begin{subfigure}[b]{\figwfactor\linewidth}
    \centering
    \includegraphics[width=\linewidth]{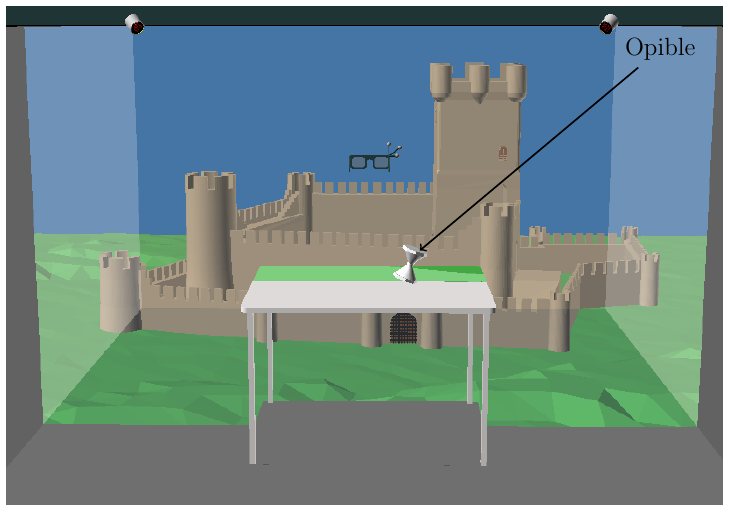}
    \caption{SABLIER: Time navigation in cultural heritage scopes (\numsablier{}).}
    \label{fig:illustration:sablier}
    \Description{A virtual reality cave displays the three-dimensional visualization of a cultural site. Viewpoint is set by tracking stereovision glasses. A table is placed at the cave's center; manipulating an hourglass (opible) enables time navigation.}
  \end{subfigure}
  \begin{subfigure}[b]{\figwfactor\linewidth}
    \centering
    \includegraphics[width=\linewidth]{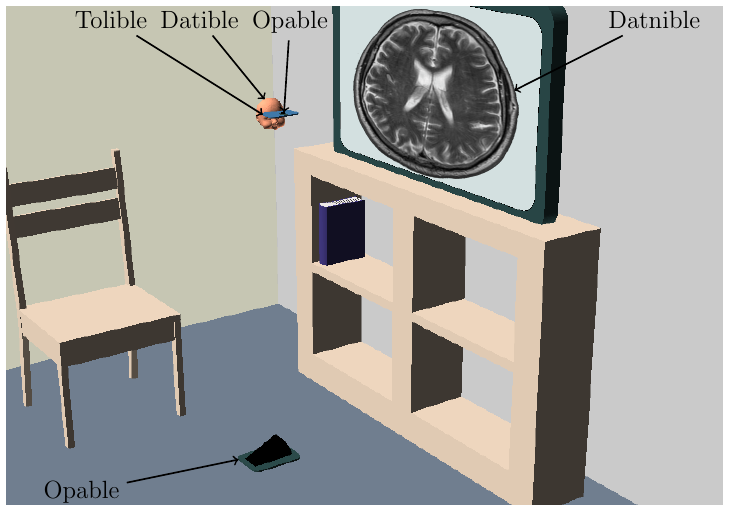}
    \caption{Head Prop: Neurosurgical visualization (\numpassiveprops{}).}
    \label{fig:illustration:passiveprops}
    \Description{A rectangular plate (tolable) is placed close to a doll's head (datible). A sizable monitor displays the resulting brain slice (datnible) from the intersection of the plate and the doll's head.}
  \end{subfigure}\hspace{\fighspace}
  \begin{subfigure}[b]{\figwfactor\linewidth}
    \centering
    \vspace{\figvspace}
    \includegraphics[width=\linewidth]{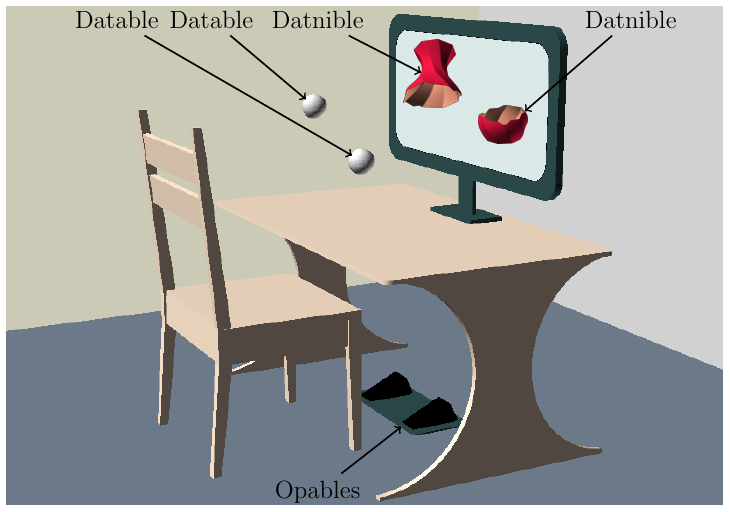}
    \caption{ArcheoTUI: Virtual archaeology fragments' assembly (\numarcheotui{}).}
    \label{fig:illustration:archeotui}
    \Description{A monitor stands on a tabletop: two 3D fragments are displayed (datnibles); two spheres (datable) held in space control the position and orientation of the two fragments; two foot-pedals (opables) stand on the ground under the tabletop. Finally, a chair enables sitting at the tabletop.}
  \end{subfigure}
  \begin{subfigure}[b]{\figwfactor\linewidth}
    \centering
    \includegraphics[width=\linewidth]{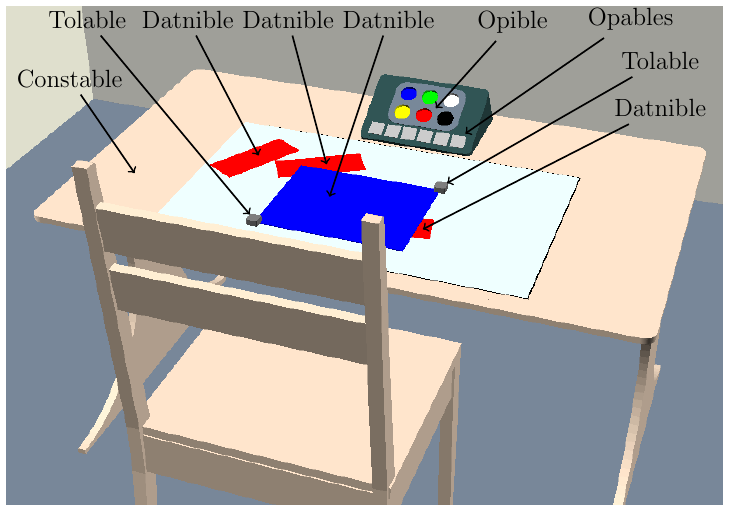}
    \caption{GraspDraw: 2D shape edition (\numgraspdraw{}).}
    \label{fig:illustration:graspdraw}
    \Description{Four rectangles (datnibles) are displayed on a tabletop: three small red rectangles and one sizable blue rectangle. The sizable rectangle is controlled by two bricks (tolables) placed on two opposite corners. A tray (constable) provides six functions in a row. An inkwell (constible) provides color picking between six colors: blue, green, white, yellow, red, and black. Finally, a chair enables sitting at the tabletop.}
  \end{subfigure}\hspace{\fighspace}
  \begin{subfigure}[b]{\figwfactor\linewidth}
    \centering
    \vspace{\figvspace}
    \includegraphics[width=\linewidth]{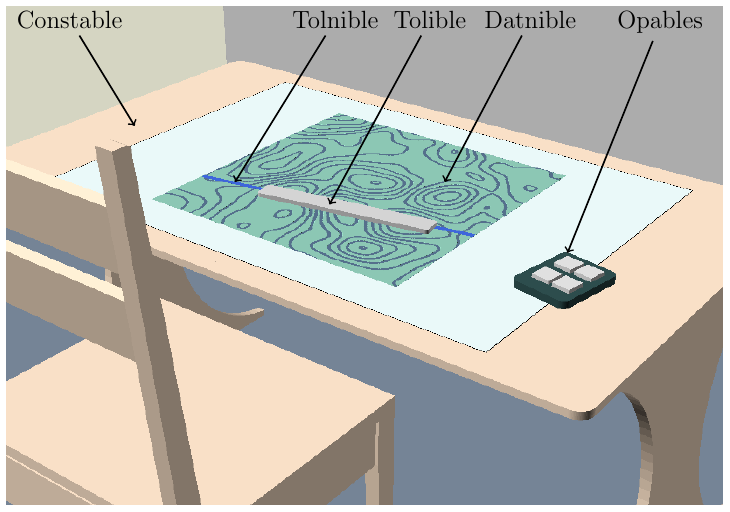}
    \caption{GeoTUI: Cutting line selection for subsoil exploration (\numgeotui{}).}
    \label{fig:illustration:geotui}
    \Description{A map (datnible) is displayed at the center of a tabletop. A ruler (tolible) controls the cutting line (tolnible) that is displayed on the map. A button box with four buttons (opables) is placed at the right of the map. Finally, a chair enables sitting at the tabletop.}
  \end{subfigure}
  \license{}\vspace{-6pt}
  \caption{Pointing out physical representations and controls for six depicted example specimen applications.}
  \label{fig:illustration}
\end{figure*}

\def\captionI{Physical representations named for \purple{14} applications from the literature, ordered by ascending years, from \purple{1976} to \purple{2001}.}

\def\captionII{Physical representations named for \purple{14} applications from the literature, ordered by ascending years, from \purple{2002} to \purple{2016}.}

\def\captionIII{Physical representations named for \purple{5} applications from the literature, ordered by ascending years, from \purple{2019} to \purple{2022}.}

\def\legend{Note. ``-'' = empty. What: `D' = Data, `T' = Tool, `O' = Operation, `C' = Constraint. How: `T' = Tangible, `G' = Graspable, `I' = Intangible. `WHT Name' = What-How Terminology name.} 

\begin{table*}
{\centering\footnotesize
  \begin{tabular}{rlllllp{0cm}llllllllll}

    \midrule

\multicolumn{4}{l}{\textbf{\textit{Application}}}
    & & \multicolumn{12}{l}{\textbf{\textit{Physical Representation}}} \tabularnewline

    \cmidrule{1-4}\cmidrule{6-17}
    
\multicolumn{4}{l}{\textbf{\textit{}}}
    & &
    & & \multicolumn{4}{l}{\textbf{\textit{What}}}
    & & \multicolumn{3}{l}{\textbf{\textit{How}}}
    & &
    \tabularnewline
    
    \cmidrule{8-11}\cmidrule{13-15}

\textit{\#}
    & \textit{Name}
    & \textit{Year}
    & \textit{References}
    & & \textit{Name or Description}
    & & \textit{D}
    & \textit{T}
    & \textit{O}
    & \textit{C}
    & & \textit{T}
    & \textit{G}
    & \textit{I}
    & & \textit{WHT Name}
    \tabularnewline
    
    \midrule
    
    1.&Slot Machine&1976&\cite{perlman1976slotmachine} and see&&Rows (procedure)&&\oo{}&\oo{}&\oo{}&\xc{}&&\oo{}&\xg{}&\oo{}&&CONSTABLE\tabularnewline
&&& in \cite[Ch.~1]{ullmer2022weaving}&&Buttons (on each row)&&\oo{}&\oo{}&\xo{}&\oo{}&&\oo{}&\xg{}&\oo{}&&OPABLE\tabularnewline
&&&&&Plastic cards (commands)&&\oo{}&\oo{}&\xo{}&\oo{}&&\oo{}&\xg{}&\oo{}&&OPABLE\tabularnewline
&&&&&Display triangle (turtle)&&\xd{}&\oo{}&\oo{}&\oo{}&&\oo{}&\oo{}&\xi{}&&DATNIBLE\tabularnewline
&&&&&Circular robot (turtle)&&\xd{}&\oo{}&\oo{}&\oo{}&&\oo{}&\xg{}&\oo{}&&DATABLE\vspace{2pt}\tabularnewline

2.&CAAD 3D Modelling System&1979&\cite{aish1979caad,aish1984caad}&&Building Blocks&&\xd{}&\oo{}&\oo{}&\oo{}&&\oo{}&\xg{}&\oo{}&&DATABLE\tabularnewline
&&&&&Design geometry (perspective view)&&\xd{}&\oo{}&\oo{}&\oo{}&&\oo{}&\oo{}&\xi{}&&DATNIBLE\tabularnewline
&&&&&Evaluation measures (Isoplots)&&\xd{}&\oo{}&\oo{}&\oo{}&&\oo{}&\oo{}&\xi{}&&DATNIBLE\vspace{2pt}\tabularnewline

3.&Self-Builder Model (Segal Model)&1980&\cite{frazer1982three,frazer1980intelligent} cited&&Panels&&\xd{}&\oo{}&\oo{}&\oo{}&&\xa{}&\oo{}&\oo{}&&DATIBLE\tabularnewline
&&&from \cite{sutphen2000reviving}&&Board&&\oo{}&\oo{}&\oo{}&\xc{}&&\oo{}&\xg{}&\oo{}&&CONSTABLE\tabularnewline
&&&&&Wireframe rendering&&\xd{}&\oo{}&\oo{}&\oo{}&&\oo{}&\oo{}&\xi{}&&DATNIBLE\tabularnewline
&&&&&Feedback tool (house area, cost)&&\xd{}&\oo{}&\oo{}&\oo{}&&\oo{}&\oo{}&\xi{}&&DATNIBLE\vspace{2pt}\tabularnewline

4.&Marble Answering Machine&1992&See in \citetext{\citealp{follmer2013inform}, \citealp{ishii1997tangible}}&&Machine&&\oo{}&\oo{}&\oo{}&\xc{}&&\oo{}&\xg{}&\oo{}&&CONSTABLE\tabularnewline
&&& and \cite[Ch.~1]{ullmer2022weaving}&&Machine's indentation (play slot)&&\oo{}&\oo{}&\xo{}&\oo{}&&\oo{}&\xg{}&\oo{}&&OPABLE\tabularnewline
&&&&&Marble (message)&&\xd{}&\oo{}&\oo{}&\oo{}&&\xa{}&\oo{}&\oo{}&&DATIBLE\tabularnewline
&&&&&Message&&\xd{}&\oo{}&\oo{}&\oo{}&&\oo{}&\oo{}&\xi{}&&DATNIBLE\vspace{2pt}\tabularnewline

5.&Head Prop&1994&\cite{hinckley1994passiveprops}&&Doll's head (brain)&&\xd{}&\oo{}&\oo{}&\oo{}&&\xa{}&\oo{}&\oo{}&&DATIBLE\tabularnewline
&&&&&Plate (slicing)&&\oo{}&\xt{}&\oo{}&\oo{}&&\xa{}&\oo{}&\oo{}&&TOLIBLE\tabularnewline
&&&&&Plate thumb button (clutch)&&\oo{}&\oo{}&\xo{}&\oo{}&&\oo{}&\xg{}&\oo{}&&OPABLE\tabularnewline
&&&&&Foot pedal (clutch)&&\oo{}&\oo{}&\xo{}&\oo{}&&\oo{}&\xg{}&\oo{}&&OPABLE\tabularnewline
&&&&&Brain 2D slice&&\xd{}&\oo{}&\oo{}&\oo{}&&\oo{}&\oo{}&\xi{}&&DATNIBLE\vspace{2pt}\tabularnewline

6.&GraspDraw&1995&\cite{fitzmaurice1995bricks}&&Bricks&&\oo{}&\xt{}&\oo{}&\oo{}&&\oo{}&\xg{}&\oo{}&&TOLABLE\tabularnewline
&&&&&2D shapes&&\xd{}&\oo{}&\oo{}&\oo{}&&\oo{}&\oo{}&\xi{}&&DATNIBLE\tabularnewline
&&&&&Inkwell&&\oo{}&\oo{}&\xo{}&\oo{}&&\xa{}&\oo{}&\oo{}&&OPIBLE\tabularnewline
&&&&&Functions' tray (select, delete, shapes)&&\oo{}&\oo{}&\xo{}&\oo{}&&\oo{}&\xg{}&\oo{}&&OPABLE\tabularnewline
&&&&&ActiveDesk surface&&\oo{}&\oo{}&\oo{}&\xc{}&&\oo{}&\xg{}&\oo{}&&CONSTABLE\vspace{2pt}\tabularnewline

7.&Tangible Geospace&1997&\cite{ishii1997tangible}&&Phicons&&\xd{}&\oo{}&\oo{}&\oo{}&&\xa{}&\oo{}&\oo{}&&DATIBLE\tabularnewline
&&&&&Campus Map&&\xd{}&\oo{}&\oo{}&\oo{}&&\oo{}&\oo{}&\xi{}&&DATNIBLE\tabularnewline
&&&&&Overlay view (passive lens)&&\oo{}&\oo{}&\xo{}&\oo{}&&\xa{}&\oo{}&\oo{}&&OPIBLE\tabularnewline
&&&&&3D view (active lens)&&\xd{}&\oo{}&\oo{}&\oo{}&&\oo{}&\oo{}&\xi{}&&DATNIBLE\tabularnewline
&&&&&Scaling and Rotating Device&&\oo{}&\xt{}&\oo{}&\oo{}&&\oo{}&\xg{}&\oo{}&&TOLABLE\tabularnewline
&&&&&Desk's surface&&\oo{}&\oo{}&\oo{}&\xc{}&&\oo{}&\xg{}&\oo{}&&CONSTABLE\vspace{2pt}\tabularnewline

8.&Build-IT&1997&\cite{rauterberg1997buildit,rauterberg1998buildit,fjeld1999camera}&&Bricks&&\oo{}&\xt{}&\oo{}&\oo{}&&\oo{}&\xg{}&\oo{}&&TOLABLE\tabularnewline
&&&&&Plan view&&\xd{}&\oo{}&\oo{}&\oo{}&&\oo{}&\oo{}&\xi{}&&DATNIBLE\tabularnewline
&&&&&Objects (robots, tables\ldots)&&\xd{}&\oo{}&\oo{}&\oo{}&&\oo{}&\oo{}&\xi{}&&DATNIBLE\tabularnewline
&&&&&Virtual cameras&&\oo{}&\oo{}&\xo{}&\oo{}&&\oo{}&\oo{}&\xi{}&&OPNIBLE\tabularnewline
&&&&&EyeCatchers&&\oo{}&\oo{}&\xo{}&\oo{}&&\oo{}&\oo{}&\xi{}&&OPNIBLE\tabularnewline
&&&&&Table surface&&\oo{}&\oo{}&\oo{}&\xc{}&&\oo{}&\xg{}&\oo{}&&CONSTABLE\tabularnewline
&&&&&3D view&&\xd{}&\oo{}&\oo{}&\oo{}&&\oo{}&\oo{}&\xi{}&&DATNIBLE\vspace{2pt}\tabularnewline

9.&Pinwheels&1998&\cite{wisneski1998pinwheels}&&Pinwheels&&\xd{}&\oo{}&\oo{}&\oo{}&&\xa{}&\oo{}&\oo{}&&DATIBLE\vspace{2pt}\tabularnewline

10.&Voodoo Dolls&1998&\cite{isidoro1998voodoo}&&Spongy object&&\xd{}&\oo{}&\oo{}&\oo{}&&\oo{}&\xg{}&\oo{}&&DATABLE\tabularnewline
&&&&&Graphics model&&\xd{}&\oo{}&\oo{}&\oo{}&&\oo{}&\oo{}&\xi{}&&DATNIBLE\vspace{2pt}\tabularnewline

11.&mediaBlocks&1998&\cite{ullmer1999mediablocks,ullmer1998mediablocks}&&mediaBlock&&\oo{}&\oo{}&\xo{}&\oo{}&&\oo{}&\xg{}&\oo{}&&OPABLE\tabularnewline
&&&&&Slots&&\oo{}&\oo{}&\oo{}&\xc{}&&\oo{}&\xg{}&\oo{}&&CONSTABLE\vspace{2pt}\tabularnewline

12.&musicBottles&1999&\cite{ishii1999musicbottles,ishii2001musicbottles}&&Music&&\xd{}&\oo{}&\oo{}&\oo{}&&\oo{}&\oo{}&\xi{}&&DATNIBLE\tabularnewline
&&&&&Bottle (music)&&\xd{}&\oo{}&\oo{}&\oo{}&&\oo{}&\xg{}&\oo{}&&DATABLE\tabularnewline
&&&&&Cork&&\oo{}&\oo{}&\xo{}&\oo{}&&\xa{}&\oo{}&\oo{}&&OPIBLE\tabularnewline
&&&&&Triangular table&&\oo{}&\oo{}&\oo{}&\xc{}&&\oo{}&\xg{}&\oo{}&&CONSTABLE\tabularnewline
&&&&&Central "stage" area&&\oo{}&\oo{}&\oo{}&\xc{}&&\oo{}&\oo{}&\xi{}&&CONSTNIBLE\vspace{2pt}\tabularnewline

13.&Urp (Urban Planning Workbench)&1999&\cite{benjoseph2001urp,ishii2002urp,underkoffler1999urp}&&Architectural Model&&\xd{}&\oo{}&\oo{}&\oo{}&&\xa{}&\oo{}&\oo{}&&DATIBLE\tabularnewline
&&&&&Road-object (strips)&&\xd{}&\oo{}&\oo{}&\oo{}&&\xa{}&\oo{}&\oo{}&&DATIBLE\tabularnewline
&&&&&Material-transformation-object (wand)&&\oo{}&\xt{}&\oo{}&\oo{}&&\oo{}&\xg{}&\oo{}&&TOLABLE\tabularnewline
&&&&&Video-camera-object&&\oo{}&\xt{}&\oo{}&\oo{}&&\xa{}&\oo{}&\oo{}&&TOLIBLE\tabularnewline
&&&&&Clock-object&&\oo{}&\xt{}&\oo{}&\oo{}&&\xa{}&\oo{}&\oo{}&&TOLIBLE\tabularnewline
&&&&&Wind-generating tool&&\oo{}&\xt{}&\oo{}&\oo{}&&\oo{}&\xg{}&\oo{}&&TOLABLE\tabularnewline
&&&&&Anemometer-object (arrow)&&\oo{}&\oo{}&\xo{}&\oo{}&&\oo{}&\xg{}&\oo{}&&OPABLE\tabularnewline
&&&&&Wind magnitude (number)&&\oo{}&\oo{}&\xo{}&\oo{}&&\oo{}&\oo{}&\xi{}&&OPNIBLE\tabularnewline
&&&&&Distance-measuring-object&&\oo{}&\oo{}&\xo{}&\oo{}&&\oo{}&\xg{}&\oo{}&&OPABLE\tabularnewline
&&&&&Distance (number)&&\oo{}&\oo{}&\xo{}&\oo{}&&\oo{}&\oo{}&\xi{}&&OPNIBLE\tabularnewline
&&&&&Shadows/Relections&&\xd{}&\oo{}&\oo{}&\oo{}&&\oo{}&\oo{}&\xi{}&&DATNIBLE\tabularnewline
&&&&&Airflow grid&&\xd{}&\oo{}&\oo{}&\oo{}&&\oo{}&\oo{}&\xi{}&&DATNIBLE\tabularnewline
&&&&&Workbench&&\oo{}&\oo{}&\oo{}&\xc{}&&\oo{}&\xg{}&\oo{}&&CONSTABLE\vspace{2pt}\tabularnewline

14.&Senseboard&2001&\cite{jacob2002senseboard}&&Grid&&\oo{}&\oo{}&\oo{}&\xc{}&&\oo{}&\oo{}&\xi{}&&CONSTNIBLE\tabularnewline
&&&&&Rectangular pucks&&\xd{}&\oo{}&\oo{}&\oo{}&&\oo{}&\xg{}&\oo{}&&DATABLE\tabularnewline
&&&&&View detail puck&&\oo{}&\xt{}&\oo{}&\oo{}&&\oo{}&\xg{}&\oo{}&&TOLABLE\tabularnewline
&&&&&Arrow puck&&\oo{}&\xt{}&\oo{}&\oo{}&&\oo{}&\xg{}&\oo{}&&TOLABLE\tabularnewline
&&&&&Values&&\xd{}&\oo{}&\oo{}&\oo{}&&\oo{}&\oo{}&\xi{}&&DATNIBLE\tabularnewline
&&&&&Board&&\oo{}&\oo{}&\oo{}&\xc{}&&\oo{}&\xg{}&\oo{}&&CONSTABLE\tabularnewline
 
    \midrule
    
  \end{tabular}\\
      \scriptsize{\textit{\legend}}
}
\caption{\captionI}~\label{tab::illustration1}
\vspace{-9pt}
\end{table*}

\begin{table*}
{\centering\footnotesize
  \begin{tabular}{rllllllllllllllll}

    \midrule

\multicolumn{4}{l}{\textbf{\textit{Application}}}
    & & \multicolumn{12}{l}{\textbf{\textit{Physical Representation}}} \tabularnewline

    \cmidrule{1-4}\cmidrule{6-17}
    
\multicolumn{4}{l}{\textbf{\textit{}}}
    & &
    & & \multicolumn{4}{l}{\textbf{\textit{What}}}
    & & \multicolumn{3}{l}{\textbf{\textit{How}}}
    & &
    \tabularnewline
    
    \cmidrule{8-11}\cmidrule{13-15}

\textit{\#}
    & \textit{Name}
    & \textit{Year}
    & \textit{References}
    & & \textit{Name or Description}
    & & \textit{D}
    & \textit{T}
    & \textit{O}
    & \textit{C}
    & & \textit{T}
    & \textit{G}
    & \textit{I}
    & & \textit{WHT Name}
    \tabularnewline
    
    \midrule

15.&Illuminating Clay&2002&\cite{ishii2004illclay,piper2002illclay}&&Clay Model&&\xd{}&\oo{}&\oo{}&\oo{}&&\xa{}&\oo{}&\oo{}&&DATIBLE\tabularnewline
&&&&&Rotative plaform&&\oo{}&\oo{}&\oo{}&\xc{}&&\oo{}&\xg{}&\oo{}&&CONSTABLE\tabularnewline
&&&&&Crosshairs&&\oo{}&\xt{}&\oo{}&\oo{}&&\oo{}&\oo{}&\xi{}&&TOLNIBLE\tabularnewline
&&&&&Cross Sections&&\xd{}&\oo{}&\oo{}&\oo{}&&\oo{}&\oo{}&\xi{}&&DATNIBLE\tabularnewline
&&&&&Analysis Function Thumbnails&&\xd{}&\oo{}&\oo{}&\oo{}&&\oo{}&\oo{}&\xi{}&&DATNIBLE\tabularnewline
&&&&&3-D perspective view&&\xd{}&\oo{}&\oo{}&\oo{}&&\oo{}&\oo{}&\xi{}&&DATNIBLE\vspace{2pt}\tabularnewline

16.&AudioPad&2002&\cite{patten2002audiopad,patten2006audiopad}&&Sounds&&\xd{}&\oo{}&\oo{}&\oo{}&&\oo{}&\oo{}&\xi{}&&DATNIBLE\tabularnewline
&&&&&Pucks (audio tracks)&&\xd{}&\oo{}&\oo{}&\oo{}&&\oo{}&\xg{}&\oo{}&&DATABLE\tabularnewline
&&&&&Star-shape puck (sound selector)&&\oo{}&\xt{}&\oo{}&\oo{}&&\oo{}&\xg{}&\oo{}&&TOLABLE\tabularnewline
&&&&&Hierarchical menu&&\oo{}&\oo{}&\xo{}&\oo{}&&\oo{}&\oo{}&\xi{}&&OPNIBLE\tabularnewline
&&&&&SenseTable surface&&\oo{}&\oo{}&\oo{}&\xc{}&&\oo{}&\xg{}&\oo{}&&CONSTABLE\vspace{2pt}\tabularnewline

17.&ReacTable&2003&\cite{jorda2007reactable}&&Music&&\xd{}&\oo{}&\oo{}&\oo{}&&\oo{}&\oo{}&\xi{}&&DATNIBLE\tabularnewline
&&&&&Point (audio output)&&\xd{}&\oo{}&\oo{}&\oo{}&&\oo{}&\oo{}&\xi{}&&DATNIBLE\tabularnewline
&&&&&Square puck (audio source)&&\xd{}&\oo{}&\oo{}&\oo{}&&\oo{}&\xg{}&\oo{}&&DATABLE\tabularnewline
&&&&&Rounded square puck (filter)&&\oo{}&\xt{}&\oo{}&\oo{}&&\oo{}&\xg{}&\oo{}&&TOLABLE\tabularnewline
&&&&&Round puck (controller)&&\oo{}&\xt{}&\oo{}&\oo{}&&\oo{}&\xg{}&\oo{}&&TOLABLE\tabularnewline
&&&&&Decagon puck (control filter)&&\oo{}&\xt{}&\oo{}&\oo{}&&\oo{}&\xg{}&\oo{}&&TOLABLE\tabularnewline
&&&&&Pentagon puck (audio mixer)&&\oo{}&\xt{}&\oo{}&\oo{}&&\oo{}&\xg{}&\oo{}&&TOLABLE\tabularnewline
&&&&&Lines (audio flow)&&\xd{}&\oo{}&\oo{}&\oo{}&&\oo{}&\oo{}&\xi{}&&DATNIBLE\tabularnewline
&&&&&Table surface&&\oo{}&\oo{}&\oo{}&\xc{}&&\oo{}&\xg{}&\oo{}&&CONSTABLE\vspace{2pt}\tabularnewline

18.&IP Network Design Workbench&2003&\cite{kobayashi2003ip}&&Pucks&&\oo{}&\xt{}&\oo{}&\oo{}&&\oo{}&\xg{}&\oo{}&&TOLABLE\tabularnewline
&&&&&Nodes&&\xd{}&\oo{}&\oo{}&\oo{}&&\oo{}&\oo{}&\xi{}&&DATNIBLE\tabularnewline
&&&&&Links&&\xd{}&\oo{}&\oo{}&\oo{}&&\oo{}&\oo{}&\xi{}&&DATNIBLE\tabularnewline
&&&&&Nodes menu&&\oo{}&\oo{}&\xo{}&\oo{}&&\oo{}&\oo{}&\xi{}&&OPNIBLE\tabularnewline
&&&&&Parameter puck (with button)&&\oo{}&\xt{}&\oo{}&\oo{}&&\oo{}&\xg{}&\oo{}&&TOLABLE\tabularnewline
&&&&&Link bandwidth&&\oo{}&\xt{}&\oo{}&\oo{}&&\oo{}&\oo{}&\xi{}&&TOLNIBLE\tabularnewline
&&&&&Router service priority&&\oo{}&\xt{}&\oo{}&\oo{}&&\oo{}&\oo{}&\xi{}&&TOLNIBLE\tabularnewline
&&&&&Number of client users&&\oo{}&\xt{}&\oo{}&\oo{}&&\oo{}&\oo{}&\xi{}&&TOLNIBLE\tabularnewline
&&&&&Server performance&&\oo{}&\xt{}&\oo{}&\oo{}&&\oo{}&\oo{}&\xi{}&&TOLNIBLE\tabularnewline
&&&&&Table surface&&\oo{}&\oo{}&\oo{}&\xc{}&&\oo{}&\xg{}&\oo{}&&CONSTABLE\tabularnewline
&&&&&Measurement graphs&&\oo{}&\oo{}&\xo{}&\oo{}&&\oo{}&\oo{}&\xi{}&&OPNIBLE\vspace{2pt}\tabularnewline

19.&Query Shapes&2004&\cite{ichida2004retrieval}&&Cubes (ActiveCubes)&&\xd{}&\oo{}&\oo{}&\oo{}&&\oo{}&\xg{}&\oo{}&&DATABLE\tabularnewline
&&&&&Voxel data representation&&\xd{}&\oo{}&\oo{}&\oo{}&&\oo{}&\oo{}&\xi{}&&DATNIBLE\tabularnewline
&&&&&3D shape models&&\xd{}&\oo{}&\oo{}&\oo{}&&\oo{}&\oo{}&\xi{}&&DATNIBLE\vspace{2pt}\tabularnewline

20.&TUISTER&2004&\cite{butz2004tuister}&&TUISTER&&\oo{}&\oo{}&\xo{}&\oo{}&&\xa{}&\oo{}&\oo{}&&OPIBLE\vspace{2pt}\tabularnewline

21.&I/O Brush&2004&\cite{ryokai2004iobrush,ryokai2007iobrush}&&Brush&&\oo{}&\xt{}&\oo{}&\oo{}&&\xa{}&\oo{}&\oo{}&&TOLIBLE\tabularnewline
&&&&&World (palette)&&\oo{}&\oo{}&\xo{}&\oo{}&&\xa{}&\oo{}&\oo{}&&OPIBLE\tabularnewline
&&&&&Canvas&&\xd{}&\oo{}&\oo{}&\oo{}&&\oo{}&\oo{}&\xi{}&&DATNIBLE\vspace{2pt}\tabularnewline

22.&PICO&2005&\cite{patten2007pico}&&Pucks (antenas)&&\xd{}&\oo{}&\oo{}&\oo{}&&\oo{}&\xg{}&\oo{}&&DATABLE\tabularnewline
&&&&&Map (city)&&\xd{}&\oo{}&\oo{}&\oo{}&&\oo{}&\oo{}&\xi{}&&DATNIBLE\tabularnewline
&&&&&Flexible curve&&\oo{}&\oo{}&\oo{}&\xc{}&&\xa{}&\oo{}&\oo{}&&CONSTIBLE\tabularnewline
&&&&&Rubber band&&\oo{}&\oo{}&\oo{}&\xc{}&&\xa{}&\oo{}&\oo{}&&CONSTIBLE\tabularnewline
&&&&&Collar&&\oo{}&\oo{}&\oo{}&\xc{}&&\xa{}&\oo{}&\oo{}&&CONSTIBLE\tabularnewline
&&&&&Teflon / Sandpaper&&\oo{}&\oo{}&\oo{}&\xc{}&&\oo{}&\xg{}&\oo{}&&CONSTABLE\tabularnewline
&&&&&Table surface&&\oo{}&\oo{}&\oo{}&\xc{}&&\oo{}&\xg{}&\oo{}&&CONSTABLE\vspace{2pt}\tabularnewline

23.&ArcheoTUI&2007&\cite{reuter2007archeotui}&&Props&&\xd{}&\oo{}&\oo{}&\oo{}&&\oo{}&\xg{}&\oo{}&&DATABLE\tabularnewline
&&&&&3D fragments&&\xd{}&\oo{}&\oo{}&\oo{}&&\oo{}&\oo{}&\xi{}&&DATNIBLE\tabularnewline
&&&&&Foot pedals (clutch)&&\oo{}&\oo{}&\xo{}&\oo{}&&\oo{}&\xg{}&\oo{}&&OPABLE\vspace{2pt}\tabularnewline

24.&GeoTUI&2008&\cite{couture2008geotui}&&Map (cube top view)&&\xd{}&\oo{}&\oo{}&\oo{}&&\oo{}&\oo{}&\xi{}&&DATNIBLE\tabularnewline
&&&&&Cutting line&&\oo{}&\xt{}&\oo{}&\oo{}&&\oo{}&\oo{}&\xi{}&&TOLNIBLE\tabularnewline
&&&&&Two-puck prop&&\oo{}&\xt{}&\oo{}&\oo{}&&\oo{}&\xg{}&\oo{}&&TOLABLE\tabularnewline
&&&&&Ruler prop&&\oo{}&\xt{}&\oo{}&\oo{}&&\xa{}&\oo{}&\oo{}&&TOLIBLE\tabularnewline
&&&&&Button box&&\oo{}&\oo{}&\xo{}&\oo{}&&\oo{}&\xg{}&\oo{}&&OPABLE\tabularnewline
&&&&&Tabletop&&\oo{}&\oo{}&\oo{}&\xc{}&&\oo{}&\xg{}&\oo{}&&CONSTABLE\vspace{2pt}\tabularnewline

25.&Slurp&2008&\cite{zigelbaum2008slurp}&&Eyedropper&&\oo{}&\oo{}&\xo{}&\oo{}&&\xa{}&\oo{}&\oo{}&&OPIBLE\vspace{2pt}\tabularnewline

26.&Relief&2010&\cite{leithinger2010relief,leithinger2011relief}&&2.5D shape display (terrain)&&\xd{}&\oo{}&\oo{}&\oo{}&&\xa{}&\oo{}&\oo{}&&DATIBLE\tabularnewline
&&&&&Topographical map&&\xd{}&\oo{}&\oo{}&\oo{}&&\oo{}&\oo{}&\xi{}&&DATNIBLE\vspace{2pt}\tabularnewline

27.&Teegi&2014&\cite{frey2014teegi}&&Teegi character (user's activity)&&\xd{}&\oo{}&\oo{}&\oo{}&&\xa{}&\oo{}&\oo{}&&DATIBLE\tabularnewline
&&&&&Brain model (user's activity)&&\xd{}&\oo{}&\oo{}&\oo{}&&\xa{}&\oo{}&\oo{}&&DATIBLE\tabularnewline
&&&&&Filter area&&\oo{}&\xt{}&\oo{}&\oo{}&&\oo{}&\oo{}&\xi{}&&TOLNIBLE\tabularnewline
&&&&&Mini-Teegis (filters)&&\oo{}&\xt{}&\oo{}&\oo{}&&\xa{}&\oo{}&\oo{}&&TOLIBLE\tabularnewline
&&&&&EEG Raw Signals&&\xd{}&\oo{}&\oo{}&\oo{}&&\oo{}&\oo{}&\xi{}&&DATNIBLE\tabularnewline
&&&&&Color map amplitude&&\oo{}&\xt{}&\oo{}&\oo{}&&\oo{}&\oo{}&\xi{}&&TOLNIBLE\tabularnewline
&&&&&Color map cursor &&\oo{}&\xt{}&\oo{}&\oo{}&&\oo{}&\xg{}&\oo{}&&TOLABLE\tabularnewline
&&&&&Table surface&&\oo{}&\oo{}&\oo{}&\xc{}&&\oo{}&\xg{}&\oo{}&&CONSTABLE\vspace{2pt}\tabularnewline

28.&SoundFORMS&2016&\cite{colter2016soundforms}&&Trigger pins&&\oo{}&\oo{}&\xo{}&\oo{}&&\oo{}&\xg{}&\oo{}&&OPABLE\tabularnewline
&&&&&Soundwave pins&&\xd{}&\oo{}&\oo{}&\oo{}&&\xa{}&\oo{}&\oo{}&&DATIBLE\tabularnewline
&&&&&Sound&&\xd{}&\oo{}&\oo{}&\oo{}&&\oo{}&\oo{}&\xi{}&&DATNIBLE\tabularnewline
 
    \midrule
    
  \end{tabular}\\
    \scriptsize{\textit{\legend}}
}
\caption{\captionII}~\label{tab::illustration2}
\vspace{-9pt}
\end{table*}

\begin{table*}
{\centering\footnotesize
  \begin{tabular}{rllllllllllllllll}

    \midrule

\multicolumn{4}{l}{\textbf{\textit{Application}}}
    & & \multicolumn{12}{l}{\textbf{\textit{Physical Representation}}} \tabularnewline

    \cmidrule{1-4}\cmidrule{6-17}
    
\multicolumn{4}{l}{\textbf{\textit{}}}
    & &
    & & \multicolumn{4}{l}{\textbf{\textit{What}}}
    & & \multicolumn{3}{l}{\textbf{\textit{How}}}
    & &
    \tabularnewline
    
    \cmidrule{8-11}\cmidrule{13-15}

\textit{\#}
    & \textit{Name}
    & \textit{Year}
    & \textit{References}
    & & \textit{Name or Description}
    & & \textit{D}
    & \textit{T}
    & \textit{O}
    & \textit{C}
    & & \textit{T}
    & \textit{G}
    & \textit{I}
    & & \textit{WHT Name}
    \tabularnewline
    
    \midrule

29.&CairnFORM&2019&\cite{daniel2019cairnform}&&Ring chart&&\xd{}&\oo{}&\oo{}&\oo{}&&\oo{}&\xg{}&\oo{}&&DATABLE\vspace{2pt}\tabularnewline

30.&reSpire&2019&\cite{choi2019respire}&&Shape-changing fabric&&\xd{}&\oo{}&\oo{}&\oo{}&&\xa{}&\oo{}&\oo{}&&DATIBLE\vspace{2pt}\tabularnewline

31.&Embodied Axes&2020&\cite{cordeil2020axes}&&Orthogonal arms (data axes)&&\oo{}&\oo{}&\xo{}&\oo{}&&\xa{}&\oo{}&\oo{}&&OPIBLE\vspace{2pt}\tabularnewline

32.&CoDa&2020&\cite{veldhuis2020coda}&&Tokens (data points)&&\xd{}&\oo{}&\oo{}&\oo{}&&\xa{}&\oo{}&\oo{}&&DATIBLE\tabularnewline
&&&&&Interactive surface&&\oo{}&\oo{}&\oo{}&\xc{}&&\oo{}&\xg{}&\oo{}&&CONSTABLE\tabularnewline
&&&&&Sidebar buttons (filter and analytic functions)&&\oo{}&\xt{}&\oo{}&\oo{}&&\oo{}&\xg{}&\oo{}&&TOLABLE\tabularnewline
&&&&&Data points&&\xd{}&\oo{}&\oo{}&\oo{}&&\oo{}&\oo{}&\xi{}&&DATNIBLE\tabularnewline
&&&&&Analytical functions (lines)&&\xd{}&\oo{}&\oo{}&\oo{}&&\oo{}&\oo{}&\xi{}&&DATNIBLE\vspace{2pt}\tabularnewline

33.&SABLIER&2022&\cite{mahieux2022sablier}&&Hourglass&&\oo{}&\oo{}&\xo{}&\oo{}&&\xa{}&\oo{}&\oo{}&&OPIBLE\tabularnewline
 
    \midrule
    
  \end{tabular}\\
    \scriptsize{\textit{\legend}}
}
\caption{\captionIII}~\label{tab::illustration3}
\vspace{-9pt}
\end{table*}

The hallmarks of the \purple{33} applicative tangible user interfaces are reported in \autoref{tab::by-classes}. This collection provides \purple{28} orthogonal hallmarks. Indeed, \purple{8} applications that share the same hallmarks are distributed into \purple{3} clusters\footnoteClusters{}. The specimens in these clusters are of various tangible user interface genres. However, some specimens of the same genre (\eg{} AudioPad (\numaudiopad{}) and ReacTable (\numreactable{})) have different hallmarks, thus demonstrating that the new holistic terminology is able to discriminate physical user interfaces at the application level (\ie{} beyond their morphological and technological descriptions). This collection also provides \purple{26} orthogonal binary hallmarks (\ie{} \purple{10} applications are now distributed into the \purple{3} clusters\footnoteClustersBinary{}).

\begin{table}
{\centering\small
  \begin{tabular}{@{\hspace{1pt}}r@{\hspace{3pt}}l@{\hspace{3pt}}l@{\hspace{3pt}}c}
\midrule
\textit{\#} & \textit{Application} & \textit{Hallmark} & \textit{Class}\tabularnewline
\midrule

1.&Slot Machine&(0, 1, 1, 0, 0, 0, 0, 2, 0, 0, 1, 0)&II\tabularnewline
2.&CAAD 3D Modelling System&(0, 1, 2, 0, 0, 0, 0, 0, 0, 0, 0, 0)&II\tabularnewline
3.&Self-Builder Model (Segal Model)&(1, 0, 2, 0, 0, 0, 0, 0, 0, 0, 1, 0)&II\tabularnewline
4.&Marble Answering Machine&(1, 0, 1, 0, 0, 0, 0, 1, 0, 0, 1, 0)&II\tabularnewline
5.&Head Prop&(1, 0, 1, 1, 0, 0, 0, 2, 0, 0, 0, 0)&II\tabularnewline
6.&GraspDraw&(0, 0, 1, 0, 1, 0, 1, 1, 0, 0, 1, 0)&III\tabularnewline
7.&Tangible Geospace&(1, 0, 2, 0, 1, 0, 1, 0, 0, 0, 1, 0)&II\tabularnewline
8.&Build-IT&(0, 0, 3, 0, 1, 0, 0, 0, 2, 0, 1, 0)&III\tabularnewline
9.&Pinwheels&(N, 0, 0, 0, 0, 0, 0, 0, 0, 0, 0, 0)&I\tabularnewline
10.&Voodoo Dolls&(0, 1, 1, 0, 0, 0, 0, 0, 0, 0, 0, 0)&II\tabularnewline
11.&mediaBlocks&(0, 0, 0, 0, 0, 0, 0, 1, 0, 0, 1, 0)&IV\tabularnewline
12.&musicBottles&(0, 1, 1, 0, 0, 0, 1, 0, 0, 0, 1, 1)&II\tabularnewline
13.&Urp (Urban Planning Workbench)&(2, 0, 2, 2, 2, 0, 0, 2, 2, 0, 1, 0)&II\tabularnewline
14.&Senseboard&(0, 1, 1, 0, 2, 0, 0, 0, 0, 0, 1, 1)&II\tabularnewline
15.&Illuminating Clay&(1, 0, 3, 0, 0, 1, 0, 0, 0, 0, 1, 0)&II\tabularnewline
16.&AudioPad&(0, 1, 1, 0, 1, 0, 0, 0, 1, 0, 1, 0)&II\tabularnewline
17.&ReacTable&(0, 1, 3, 0, 4, 0, 0, 0, 0, 0, 1, 0)&II\tabularnewline
18.&IP Network Design Workbench&(0, 0, 2, 0, 2, 4, 0, 0, 2, 0, 1, 0)&III\tabularnewline
19.&Query Shapes&(0, 1, 2, 0, 0, 0, 0, 0, 0, 0, 0, 0)&II\tabularnewline
20.&TUISTER&(0, 0, 0, 0, 0, 0, 1, 0, 0, 0, 0, 0)&IV\tabularnewline
21.&I/O Brush&(0, 0, 1, 1, 0, 0, 1, 0, 0, 0, 0, 0)&III\tabularnewline
22.&PICO&(0, 1, 1, 0, 0, 0, 0, 0, 0, 3, 2, 0)&II\tabularnewline
24.&GeoTUI&(0, 0, 1, 1, 1, 1, 0, 1, 0, 0, 1, 0)&III\tabularnewline
25.&Slurp&(0, 0, 0, 0, 0, 0, 1, 0, 0, 0, 0, 0)&IV\tabularnewline
26.&Relief&(1, 0, 0, 0, 0, 0, 0, 0, 0, 0, 0, 0)&I\tabularnewline
27.&Teegi&(2, 0, 1, 1, 1, 2, 0, 0, 0, 0, 1, 0)&II\tabularnewline
28.&SoundFORMS&(1, 0, 1, 0, 0, 0, 0, 1, 0, 0, 0, 0)&II\tabularnewline
29.&CairnFORM&(0, 1, 0, 0, 0, 0, 0, 0, 0, 0, 0, 0)&I\tabularnewline
30.&reSpire&(1, 0, 0, 0, 0, 0, 0, 0, 0, 0, 0, 0)&I\tabularnewline
31.&Embodied Axes&(0, 0, 0, 0, 0, 0, 1, 0, 0, 0, 0, 0)&IV\tabularnewline
32.&CoDa&(1, 0, 2, 0, 1, 0, 0, 0, 0, 0, 1, 0)&II\tabularnewline
33.&SABLIER&(0, 0, 0, 0, 0, 0, 1, 0, 0, 0, 0, 0)&IV\tabularnewline
 
\midrule
  \end{tabular}
}
\caption{Hallmarks and classes of the \purple{33} applications.}~\label{tab::by-classes}
\vspace{-9pt}
\end{table}

\subsection{Corpus Classification}

The classification of the corpus through the four classes is shown in \autoref{tab::by-classes}: the collection comprises \purple{4} applications of Class~I, \purple{19} of Class~II, \purple{5} of Class~III, and \purple{5} of Class~IV. All of the applications could be distributed among only the four classes. Only a few applications of the corpus represented data through only tangible or graspable representation (\ie{} \purple{4} applications of Class~I), whereas over half combined them with intangible data representations (\ie{} \purple{19} applications of Class~II). Moreover, only a few applications merely combine intangible data representations with tangible or graspable tools (\purple{5} applications of Class~III). However, this distribution may result from the corpus's constitution (\ie{} by selecting orthogonal hallmarks), and may differ in a more extended corpus (\eg{} by gathering applications through a systematic literature review).

The collection shows several possible configurations for Class~I applications, from duplication of the same datibles, to datables combined with tolables and opnibles, to a unique datible. For example, the ambient interface Pinwheels (\numpinwheels{}) uses multiple replicated datibles: the \g{pinwheels} rotation speed then informs about a flow (\eg{} stock market exchange, car traffic, or e-mail traffic).
CairnFORM (\numcairnform{}) is one datible made of replicated rings, whose expansion informs about renewable energy forecasts.
Finally, Relief (\numrelief{}) and reSpire (\numrespire{}) are single deformable datibles, that change surfaces' shapes to modelize terrain topography and echo self-breathing, respectively.

The collection includes cases for Class~II applications where datnibles are bound or combined with datibles or datables. For example, the Head Prop specimen (\numpassiveprops{}) binds the intersection between a datible (\g{doll's head}) and a tolible (\g{plate}) to a datnible (\g{slice view}). On the tabletop ReacTable (\numreactable{}), audio sources are represented by datables (\g{square pucks}), audio flow and audio output are materialized by datnibles (some \g{lines} and a \g{dot,} respectively); some tolables allow mixing or filtering the audio flow. Finally, the actuated interface SoundFORMS (\numsoundform{}) assigns digital roles to an array of actuated pins: pressing some opables (\g{trigger pins}) starts playing datnibles (\g{sounds}) that are bound to datibles (\g{soundwave pins}).

In Class~III applications, the tangible or graspable part of the user interface is not about data, but rather tools or operations. For example, the I/O Brush (\numiobrush{}) uses World as an opible (\g{palette}), to draw with a tolible (\g{brush}) on a datnible (\g{canvas}). The tabletop GraspDraw (\numgraspdraw{}) edits datnibles (\g{shapes}) using a tolable (\g{pucks}) that serve as physical anchors. Finally, the tabletop GeoTUI (\numgeotui{}) controls a tolnible (\g{cutting line}) with a tolible (\g{ruler}) or a tolable (\g{two-puck prop}) on a datnible (\g{top view map}).

Artifacts of Class~IV are designed to integrate across various applications, whose scopes can extend beyond tangible user interfaces (\eg{} graphical user interfaces \cite{zigelbaum2008slurp}, augmented reality \cite{cordeil2020axes}, and virtual reality caves \cite{mahieux2022sablier}). For example, Slurp (\numslurp{}) is an \g{eyedropper} that can extract and inject digital information. TUISTER (\numtuister{}), Embodied Axes (\numaxes{}), and SABLIER (\numsablier{}) provide artifacts to browse hierarchical structures, control 3D positions, and navigate through space and time, respectively.

\section{Mapping the Previous Terms}

\autoref{fig:conceptual-space} illustrates a mapping between the previous terms found in the literature and the new holistic terminology, achieved through a projection onto the two axes of the taxonomy that originated the terminology (\ie{} ``what'' is physically represented and ``how''). Only one term can cover the whole terminology. Except for this holistic term, the mapping reveals a focus of previous research on representing tangible data, which is describable through half of the terms. In contrast, constraints and intangible data are describable by only one term each, and intangible tools and operations are outside the conceptual perception.

\def\txtTAr{\BT{Tangible representations \cite{ullmer2002phdthesis,ullmer2005token}}}
\def\txtINr{\BT{Intangible representations \cite{ullmer2002phdthesis,ullmer2005token}}}
\def\txtCOR{\BT{Core Tangibles \cite{ullmer2008core}}}

\def\txtTAN{\WT{Tangibles\,\cite{ullmer2000emerging}}}
\def\txtART{\WT{Artifacts \cite{ullmer2000emerging}}}
\def\txtPHIC{\WT{Phicons \cite{ullmer1997metadesk}}}
\def\txtPASS{\WT{Passive Props \cite{hinckley1994passiveprops}}}
\def\txtPHAN{\WT{Phandles \cite{ullmer1997metadesk}}}
\def\txtPHID{\WT{Phidgets \cite{greenberg2001phidgets}}}

\def\txtOBJ{\BT{R\textsubscript{object} \cite{dubois2003asur++}}}
\def\txtTOL{\BT{R\textsubscript{tool} \cite{dubois2003asur++}}}

\def\txtTOK{\WT{Tokens}\\ \WT{\cite{shaer2004tac,ullmer2002phdthesis,ullmer2005token}}}
\def\txtCON{\WT{Constraints}\\ \WT{\hspace{1pt}\cite{shaer2004tac,ullmer2002phdthesis,ullmer2005token}}}
\def\txtPYF{\BT{Pyfos \cite{shaer2004tac}}}

\def\legEX{\BT{Examples of Media}}
\def\legGEN{\BT{General Terms}}
\def\legISO{\BT{Isolated Terms}}
\def\legASUR{\BT{ASUR++ Terms \cite{dubois2003asur++}}}
\def\legTAC{\BT{TAC Terms \cite{shaer2004tac}}}

\begin{figure*}[h]
  \centering
  \def\svgwidth{\textwidth} {\small \begingroup \makeatletter \providecommand\color[2][]{\errmessage{(Inkscape) Color is used for the text in Inkscape, but the package 'color.sty' is not loaded}\renewcommand\color[2][]{}}\providecommand\transparent[1]{\errmessage{(Inkscape) Transparency is used (non-zero) for the text in Inkscape, but the package 'transparent.sty' is not loaded}\renewcommand\transparent[1]{}}\providecommand\rotatebox[2]{#2}\newcommand*\fsize{\dimexpr\f@size pt\relax}\newcommand*\lineheight[1]{\fontsize{\fsize}{#1\fsize}\selectfont}\ifx\svgwidth\undefined \setlength{\unitlength}{768.0167419bp}\ifx\svgscale\undefined \relax \else \setlength{\unitlength}{\unitlength * \real{\svgscale}}\fi \else \setlength{\unitlength}{\svgwidth}\fi \global\let\svgwidth\undefined \global\let\svgscale\undefined \makeatother \begin{picture}(1,0.54149305)\lineheight{1}\setlength\tabcolsep{0pt}\put(0.56536042,0.00491629){\makebox(0,0)[lt]{\lineheight{1.25}\smash{\begin{tabular}[t]{l}\textit{\legASUR{}}\end{tabular}}}}\put(0,0){\includegraphics[width=\unitlength,page=1]{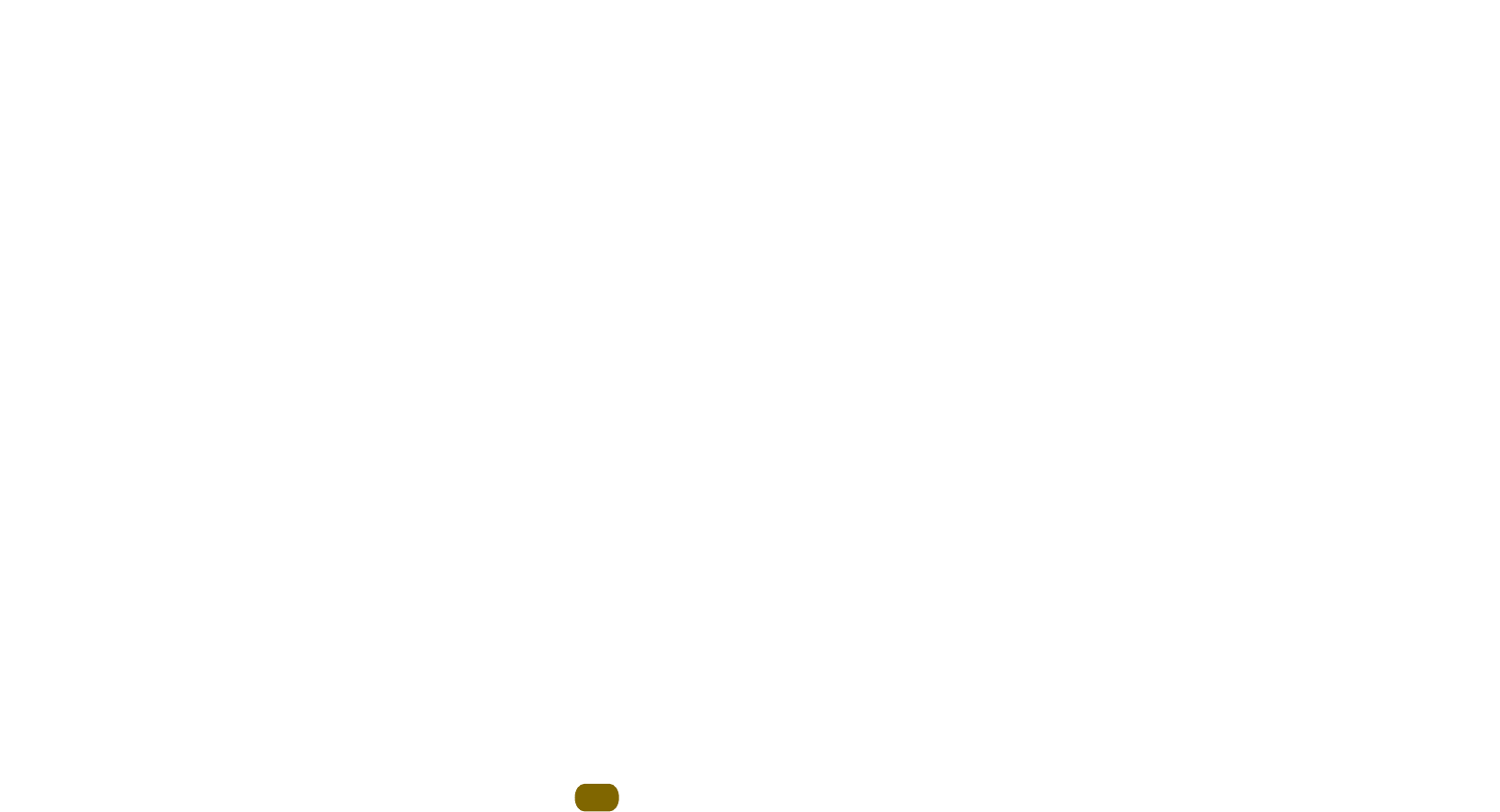}}\put(0.11922173,0.00491655){\makebox(0,0)[lt]{\lineheight{1.25}\smash{\begin{tabular}[t]{l}\textit{\legGEN{}}\end{tabular}}}}\put(0,0){\includegraphics[width=\unitlength,page=2]{conceptual-space6.pdf}}\put(0.25231374,0.00491655){\makebox(0,0)[lt]{\lineheight{1.25}\smash{\begin{tabular}[t]{l}\textit{\legEX{}}\end{tabular}}}}\put(0,0){\includegraphics[width=\unitlength,page=3]{conceptual-space6.pdf}}\put(0.0081783,0.00491655){\makebox(0,0)[lt]{\lineheight{1.25}\smash{\begin{tabular}[t]{l}\BT{Legend:}\end{tabular}}}}\put(0.74062431,0.00491655){\makebox(0,0)[lt]{\lineheight{1.25}\smash{\begin{tabular}[t]{l}\textit{\legTAC{}}\end{tabular}}}}\put(0.42032099,0.00491655){\makebox(0,0)[lt]{\lineheight{1.25}\smash{\begin{tabular}[t]{l}\textit{\legISO{}}\end{tabular}}}}\put(0,0){\includegraphics[width=\unitlength,page=4]{conceptual-space6.pdf}}\put(0.05784689,0.40182705){\makebox(0,0)[t]{\lineheight{1.25}\smash{\begin{tabular}[t]{c}TANGIBLE\end{tabular}}}}\put(0.05738532,0.25087068){\makebox(0,0)[t]{\lineheight{1.25}\smash{\begin{tabular}[t]{c}GRASPABLE\end{tabular}}}}\put(0.05705727,0.09991452){\makebox(0,0)[t]{\lineheight{1.25}\smash{\begin{tabular}[t]{c}INTANGIBLE\end{tabular}}}}\put(0.44582521,0.4906085){\makebox(0,0)[lt]{\lineheight{1.25}\smash{\begin{tabular}[t]{l}TOOL\end{tabular}}}}\put(0.65524514,0.4906085){\makebox(0,0)[lt]{\lineheight{1.25}\smash{\begin{tabular}[t]{l}OPERATION\end{tabular}}}}\put(0.85244056,0.4906085){\makebox(0,0)[lt]{\lineheight{1.25}\smash{\begin{tabular}[t]{l}CONSTRAINT\end{tabular}}}}\put(0,0){\includegraphics[width=\unitlength,page=5]{conceptual-space6.pdf}}\put(0.05685128,0.49072868){\makebox(0,0)[t]{\lineheight{1.25}\smash{\begin{tabular}[t]{c}\textbf{HOW}\end{tabular}}}}\put(0.53016359,0.52103433){\makebox(0,0)[lt]{\lineheight{1.25}\smash{\begin{tabular}[t]{l}\textbf{WHAT}\end{tabular}}}}\put(0,0){\includegraphics[width=\unitlength,page=6]{conceptual-space6.pdf}}\put(0.70572109,0.42532126){\makebox(0,0)[t]{\lineheight{0.94999999}\smash{\begin{tabular}[t]{c}\txtCOR{}\end{tabular}}}}\put(0,0){\includegraphics[width=\unitlength,page=7]{conceptual-space6.pdf}}\put(0.89370085,0.00491655){\color[rgb]{0,0,0}\makebox(0,0)[lt]{\lineheight{1.25}\smash{\begin{tabular}[t]{l}\txtPYF{}\end{tabular}}}}\put(0.90770907,0.29579779){\color[rgb]{1,1,1}\makebox(0,0)[t]{\lineheight{0.89999998}\smash{\begin{tabular}[t]{c}\txtCON{}\end{tabular}}}}\put(0.2034415,0.49072868){\makebox(0,0)[lt]{\lineheight{1.25}\smash{\begin{tabular}[t]{l}DATUM\end{tabular}}}}\put(0,0){\includegraphics[width=\unitlength,page=8]{conceptual-space6.pdf}}\put(0.13866443,0.15234582){\makebox(0,0)[lt]{\lineheight{0.89999998}\smash{\begin{tabular}[t]{l}\txtINr{}\end{tabular}}}}\put(0,0){\includegraphics[width=\unitlength,page=9]{conceptual-space6.pdf}}\put(0.13866443,0.44472913){\makebox(0,0)[lt]{\lineheight{0.89999998}\smash{\begin{tabular}[t]{l}\txtTAr{}\end{tabular}}}}\put(0.27233338,0.31981175){\makebox(0,0)[rt]{\lineheight{1.25}\smash{\begin{tabular}[t]{r}\txtOBJ{}\end{tabular}}}}\put(0,0){\includegraphics[width=\unitlength,page=10]{conceptual-space6.pdf}}\put(0.17485365,0.08005195){\color[rgb]{0,0,0}\makebox(0,0)[lt]{\lineheight{1.25}\smash{\begin{tabular}[t]{l}\BT{Sound}\end{tabular}}}}\put(0.49888681,0.42456575){\makebox(0,0)[rt]{\lineheight{1.25}\smash{\begin{tabular}[t]{r}\txtTOL{}\end{tabular}}}}\put(0,0){\includegraphics[width=\unitlength,page=11]{conceptual-space6.pdf}}\put(0.2843097,0.08005195){\color[rgb]{0,0,0}\makebox(0,0)[t]{\lineheight{1.25}\smash{\begin{tabular}[t]{c}\BT{Video}\end{tabular}}}}\put(0,0){\includegraphics[width=\unitlength,page=12]{conceptual-space6.pdf}}\put(0.1565538,0.35252036){\makebox(0,0)[lt]{\lineheight{0.89999998}\smash{\begin{tabular}[t]{l}\txtPASS{}\end{tabular}}}}\put(0,0){\includegraphics[width=\unitlength,page=13]{conceptual-space6.pdf}}\put(0.1565538,0.37660311){\makebox(0,0)[lt]{\lineheight{0.89999998}\smash{\begin{tabular}[t]{l}\txtPHIC{}\end{tabular}}}}\put(0.1565538,0.40077829){\makebox(0,0)[lt]{\lineheight{0.89999998}\smash{\begin{tabular}[t]{l}\txtART{}\end{tabular}}}}\put(0.1565538,0.42447673){\makebox(0,0)[lt]{\lineheight{0.89999998}\smash{\begin{tabular}[t]{l}\txtTAN{}\end{tabular}}}}\put(0.4646626,0.33640355){\makebox(0,0)[t]{\lineheight{0.94999999}\smash{\begin{tabular}[t]{c}\txtTOK{}\end{tabular}}}}\put(0,0){\includegraphics[width=\unitlength,page=14]{conceptual-space6.pdf}}\put(0.42480169,0.23929891){\makebox(0,0)[lt]{\lineheight{0.89999998}\smash{\begin{tabular}[t]{l}\txtPHAN{}\end{tabular}}}}\put(0.66233015,0.24320508){\makebox(0,0)[lt]{\lineheight{0.89999998}\smash{\begin{tabular}[t]{l}\txtPHID{}\end{tabular}}}}\end{picture}\endgroup  }
  \caption{Mapping \purple{thirteen} previous terms for physical representations and controls on the new conceptual space.}
  \Description{The infography is a graphical projection of the previous terms on a two-dimensional space determined by the what--how taxonomy that originated the terms of the new terminology. The horizontal axis is composed of the four values about what digital role is assigned: Datum, Tool, Operation, or Constraint. The vertical axis is composed of three tangibility degrees: Tangible, Graspable, and Intangible. The resulting grid is covered by rectangles according to previous terms mapping on the two axes.}
  \label{fig:conceptual-space}
\end{figure*}
 
The distorted focus on naming tangible data may have its origins in the community's search for paradigms and frameworks, as well as the influence of the data-centered viewpoint in the definition of tangible user interfaces. However, other roles than data are required to describe applicative tangible user interfaces.

The following section proposes to match the four tangibility classes with some previous classifications through the set of (orthogonal) applications provided by the corpus.

\section{Cross-Matching Classifications}

\begin{figure*}[h]
  \centering
\def\svgwidth{\textwidth} {\scriptsize \begingroup \makeatletter \providecommand\color[2][]{\errmessage{(Inkscape) Color is used for the text in Inkscape, but the package 'color.sty' is not loaded}\renewcommand\color[2][]{}}\providecommand\transparent[1]{\errmessage{(Inkscape) Transparency is used (non-zero) for the text in Inkscape, but the package 'transparent.sty' is not loaded}\renewcommand\transparent[1]{}}\providecommand\rotatebox[2]{#2}\newcommand*\fsize{\dimexpr\f@size pt\relax}\newcommand*\lineheight[1]{\fontsize{\fsize}{#1\fsize}\selectfont}\ifx\svgwidth\undefined \setlength{\unitlength}{847.29261503bp}\ifx\svgscale\undefined \relax \else \setlength{\unitlength}{\unitlength * \real{\svgscale}}\fi \else \setlength{\unitlength}{\svgwidth}\fi \global\let\svgwidth\undefined \global\let\svgscale\undefined \makeatother \begin{picture}(1,1.0210417)\lineheight{1}\setlength\tabcolsep{0pt}\put(0.38310133,1.01186448){\color[rgb]{0,0,0}\makebox(0,0)[t]{\lineheight{1.25}\smash{\begin{tabular}[t]{c}\textbf{\textit{Subgenre}}\end{tabular}}}}\put(0.09098118,1.01186448){\color[rgb]{0,0,0}\makebox(0,0)[t]{\lineheight{1.25}\smash{\begin{tabular}[t]{c}\textbf{\textit{Genre}}\end{tabular}}}}\put(0.93893216,1.01186448){\color[rgb]{0,0,0}\makebox(0,0)[t]{\lineheight{1.25}\smash{\begin{tabular}[t]{c}\textbf{\textit{Tangibility Class}}\end{tabular}}}}\put(0.70602575,1.01207194){\color[rgb]{0,0,0}\makebox(0,0)[t]{\lineheight{1.25}\smash{\begin{tabular}[t]{c}\textbf{\textit{Application}}\end{tabular}}}}\put(0,0){\includegraphics[width=\unitlength,page=1]{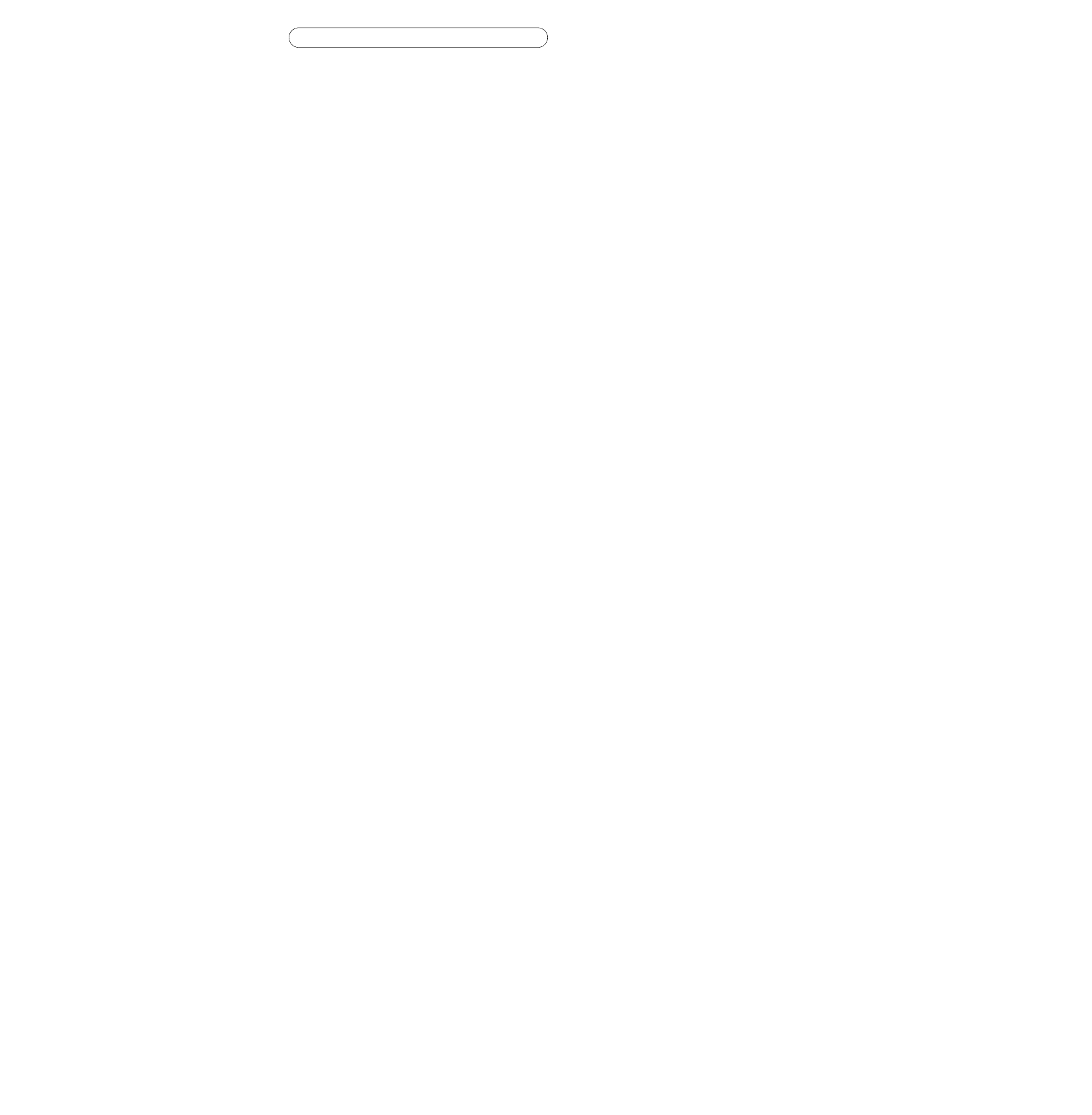}}\put(0.38242198,0.98221194){\color[rgb]{0,0,0}\makebox(0,0)[t]{\lineheight{1.25}\smash{\begin{tabular}[t]{c}Dynamic everyday objects\end{tabular}}}}\put(0.70577511,0.98290636){\color[rgb]{0,0,0.06666667}\makebox(0,0)[t]{\lineheight{1.25}\smash{\begin{tabular}[t]{c}(\#9) Pinwheels\end{tabular}}}}\put(0,0){\includegraphics[width=\unitlength,page=2]{cross-classification.pdf}}\put(0.09136159,0.98221194){\color[rgb]{0,0,0}\makebox(0,0)[t]{\lineheight{1.25}\smash{\begin{tabular}[t]{c}Ambient Media\end{tabular}}}}\put(0,0){\includegraphics[width=\unitlength,page=3]{cross-classification.pdf}}\put(0.93892931,0.98132859){\color[rgb]{0,0,0}\makebox(0,0)[t]{\lineheight{1.25}\smash{\begin{tabular}[t]{c}Class I\end{tabular}}}}\put(0,0){\includegraphics[width=\unitlength,page=4]{cross-classification.pdf}}\put(0.09125229,0.05087828){\color[rgb]{0,0,0}\makebox(0,0)[t]{\lineheight{1.25}\smash{\begin{tabular}[t]{c}Tokens and\\Constraints\end{tabular}}}}\put(0,0){\includegraphics[width=\unitlength,page=5]{cross-classification.pdf}}\put(0.0911314,0.26327595){\color[rgb]{0,0,0}\makebox(0,0)[t]{\lineheight{1.25}\smash{\begin{tabular}[t]{c}Interactive Surfaces\\and Spaces\end{tabular}}}}\put(0,0){\includegraphics[width=\unitlength,page=6]{cross-classification.pdf}}\put(0.70537458,0.34651074){\color[rgb]{0,0,0.06666667}\makebox(0,0)[t]{\lineheight{1.25}\smash{\begin{tabular}[t]{c}(\#22) PICO\end{tabular}}}}\put(0.70537458,0.40699529){\color[rgb]{0,0,0.06666667}\makebox(0,0)[t]{\lineheight{1.25}\smash{\begin{tabular}[t]{c}(\#14) Senseboard\end{tabular}}}}\put(0.70537458,0.31626556){\color[rgb]{0,0,0.06666667}\makebox(0,0)[t]{\lineheight{1.25}\smash{\begin{tabular}[t]{c}(\#7) Tangible Geospace\end{tabular}}}}\put(0.70537458,0.28601461){\color[rgb]{0,0,0.06666667}\makebox(0,0)[t]{\lineheight{1.25}\smash{\begin{tabular}[t]{c}(\#16) AudioPad\end{tabular}}}}\put(0.70537458,0.25576948){\color[rgb]{0,0,0.06666667}\makebox(0,0)[t]{\lineheight{1.25}\smash{\begin{tabular}[t]{c}(\#17) ReacTable\end{tabular}}}}\put(0.70537458,0.22553012){\color[rgb]{0,0,0.06666667}\makebox(0,0)[t]{\lineheight{1.25}\smash{\begin{tabular}[t]{c}(\#32) CoDa\end{tabular}}}}\put(0.70537458,0.19573442){\color[rgb]{0,0,0.06666667}\makebox(0,0)[t]{\lineheight{1.25}\smash{\begin{tabular}[t]{c}(\#6) GraspDraw\end{tabular}}}}\put(0.70537458,0.16503404){\color[rgb]{0,0,0.06666667}\makebox(0,0)[t]{\lineheight{1.25}\smash{\begin{tabular}[t]{c}(\#8) Build-IT\end{tabular}}}}\put(0.70537458,0.13523763){\color[rgb]{0,0,0.06666667}\makebox(0,0)[t]{\lineheight{1.25}\smash{\begin{tabular}[t]{c}(\#18) IP Network Design Workbench\end{tabular}}}}\put(0.70537458,0.10454873){\color[rgb]{0,0,0.06666667}\makebox(0,0)[t]{\lineheight{1.25}\smash{\begin{tabular}[t]{c}(\#24) GeoTUI\end{tabular}}}}\put(0,0){\includegraphics[width=\unitlength,page=7]{cross-classification.pdf}}\put(0.38245649,0.34695446){\color[rgb]{0,0,0}\makebox(0,0)[t]{\lineheight{1.25}\smash{\begin{tabular}[t]{c}Actuated tabletop\end{tabular}}}}\put(0.38213672,0.40630082){\color[rgb]{0,0,0}\makebox(0,0)[t]{\lineheight{1.25}\smash{\begin{tabular}[t]{c}Board\end{tabular}}}}\put(0.38251997,0.21085125){\color[rgb]{0,0,0}\makebox(0,0)[t]{\lineheight{1.25}\smash{\begin{tabular}[t]{c}Tabletop\end{tabular}}}}\put(0.38248528,0.37605569){\color[rgb]{0,0,0}\makebox(0,0)[t]{\lineheight{1.25}\smash{\begin{tabular}[t]{c}Workbench\end{tabular}}}}\put(0.70537458,0.37720541){\color[rgb]{0,0,0.06666667}\makebox(0,0)[t]{\lineheight{1.25}\smash{\begin{tabular}[t]{c}(\#13) Urp\end{tabular}}}}\put(0.70537458,0.07429773){\color[rgb]{0,0,0.06666667}\makebox(0,0)[t]{\lineheight{1.25}\smash{\begin{tabular}[t]{c}(\#1) Slot Machine\end{tabular}}}}\put(0.70537458,0.04450209){\color[rgb]{0,0,0.06666667}\makebox(0,0)[t]{\lineheight{1.25}\smash{\begin{tabular}[t]{c}(\#4) Marble Answering Machine\end{tabular}}}}\put(0.70484449,0.01380747){\color[rgb]{0,0,0.06666667}\makebox(0,0)[t]{\lineheight{1.25}\smash{\begin{tabular}[t]{c}(\#11) mediaBlocks\end{tabular}}}}\put(0,0){\includegraphics[width=\unitlength,page=8]{cross-classification.pdf}}\put(0.93892931,0.34528971){\color[rgb]{0,0,0}\makebox(0,0)[t]{\lineheight{1.25}\smash{\begin{tabular}[t]{c}Class II\end{tabular}}}}\put(0,0){\includegraphics[width=\unitlength,page=9]{cross-classification.pdf}}\put(0.93892931,0.15269274){\color[rgb]{0,0,0}\makebox(0,0)[t]{\lineheight{1.25}\smash{\begin{tabular}[t]{c}Class III\end{tabular}}}}\put(0,0){\includegraphics[width=\unitlength,page=10]{cross-classification.pdf}}\put(0.93892931,0.05773546){\color[rgb]{0,0,0}\makebox(0,0)[t]{\lineheight{1.25}\smash{\begin{tabular}[t]{c}Class II\end{tabular}}}}\put(0,0){\includegraphics[width=\unitlength,page=11]{cross-classification.pdf}}\put(0.93892931,0.013113){\color[rgb]{0,0,0}\makebox(0,0)[t]{\lineheight{1.25}\smash{\begin{tabular}[t]{c}Class IV\end{tabular}}}}\put(0,0){\includegraphics[width=\unitlength,page=12]{cross-classification.pdf}}\put(0.09136159,0.73786063){\color[rgb]{0,0,0}\makebox(0,0)[t]{\lineheight{1.25}\smash{\begin{tabular}[t]{c}Artifacts \& Objects\end{tabular}}}}\put(0,0){\includegraphics[width=\unitlength,page=13]{cross-classification.pdf}}\put(0.0910605,0.47282438){\color[rgb]{0,0,0}\makebox(0,0)[t]{\lineheight{1.25}\smash{\begin{tabular}[t]{c}Constructive\\Assemblies\end{tabular}}}}\put(0.3818888,0.46536294){\color[rgb]{0,0,0}\makebox(0,0)[t]{\lineheight{1.25}\smash{\begin{tabular}[t]{c}Mirrored constructive assemblies\end{tabular}}}}\put(0,0){\includegraphics[width=\unitlength,page=14]{cross-classification.pdf}}\put(0.70537458,0.49771919){\color[rgb]{0,0,0.06666667}\makebox(0,0)[t]{\lineheight{1.25}\smash{\begin{tabular}[t]{c}(\#2) CAAD 3D Modelling System\end{tabular}}}}\put(0.70537458,0.46702458){\color[rgb]{0,0,0.06666667}\makebox(0,0)[t]{\lineheight{1.25}\smash{\begin{tabular}[t]{c}(\#3) Self-Builder Model\end{tabular}}}}\put(0.70537458,0.43722893){\color[rgb]{0,0,0.06666667}\makebox(0,0)[t]{\lineheight{1.25}\smash{\begin{tabular}[t]{c}(\#19) Query Shapes\end{tabular}}}}\put(0.70537458,0.64849513){\color[rgb]{0,0,0.06666667}\makebox(0,0)[t]{\lineheight{1.25}\smash{\begin{tabular}[t]{c}(\#21) I/O Brush\end{tabular}}}}\put(0.70484449,0.55821493){\color[rgb]{0,0,0.06666667}\makebox(0,0)[t]{\lineheight{1.25}\smash{\begin{tabular}[t]{c}(\#20) TUISTER\end{tabular}}}}\put(0.70484449,0.527964){\color[rgb]{0,0,0.06666667}\makebox(0,0)[t]{\lineheight{1.25}\smash{\begin{tabular}[t]{c}(\#31) Embodied Axes\end{tabular}}}}\put(0.70484449,0.61825587){\color[rgb]{0,0,0.06666667}\makebox(0,0)[t]{\lineheight{1.25}\smash{\begin{tabular}[t]{c}(\#33) SABLIER\end{tabular}}}}\put(0,0){\includegraphics[width=\unitlength,page=15]{cross-classification.pdf}}\put(0.93892931,0.76811525){\color[rgb]{0,0,0}\makebox(0,0)[t]{\lineheight{1.25}\smash{\begin{tabular}[t]{c}Class II\end{tabular}}}}\put(0,0){\includegraphics[width=\unitlength,page=16]{cross-classification.pdf}}\put(0.93892931,0.64764264){\color[rgb]{0,0,0}\makebox(0,0)[t]{\lineheight{1.25}\smash{\begin{tabular}[t]{c}Class III\end{tabular}}}}\put(0,0){\includegraphics[width=\unitlength,page=17]{cross-classification.pdf}}\put(0.93892931,0.46580931){\color[rgb]{0,0,0}\makebox(0,0)[t]{\lineheight{1.25}\smash{\begin{tabular}[t]{c}Class II\end{tabular}}}}\put(0,0){\includegraphics[width=\unitlength,page=18]{cross-classification.pdf}}\put(0.93892931,0.57081748){\color[rgb]{0,0,0}\makebox(0,0)[t]{\lineheight{1.25}\smash{\begin{tabular}[t]{c}Class IV\end{tabular}}}}\put(0,0){\includegraphics[width=\unitlength,page=19]{cross-classification.pdf}}\put(0.38188881,0.04450209){\color[rgb]{0,0,0}\makebox(0,0)[t]{\lineheight{1.25}\smash{\begin{tabular}[t]{c}Symbolic tokens \cite{ullmer2000emerging}\end{tabular}}}}\put(0,0){\includegraphics[width=\unitlength,page=20]{cross-classification.pdf}}\put(0.93892931,0.91990969){\color[rgb]{0,0,0}\makebox(0,0)[t]{\lineheight{1.25}\smash{\begin{tabular}[t]{c}Class I\end{tabular}}}}\put(0,0){\includegraphics[width=\unitlength,page=21]{cross-classification.pdf}}\put(0.38354508,0.90674067){\color[rgb]{0,0,0}\makebox(0,0)[t]{\lineheight{1.25}\smash{\begin{tabular}[t]{c}Transformable continuous tangibles \cite{ishii2012radical}\end{tabular}}}}\put(0.70537458,0.92286593){\color[rgb]{0,0,0.05882353}\makebox(0,0)[t]{\lineheight{1.25}\smash{\begin{tabular}[t]{c}(\#29) reSpire\end{tabular}}}}\put(0.70563676,0.89031327){\color[rgb]{0,0,0.06666667}\makebox(0,0)[t]{\lineheight{1.25}\smash{\begin{tabular}[t]{c}(\#30) CairnFORM\end{tabular}}}}\put(0,0){\includegraphics[width=\unitlength,page=22]{cross-classification.pdf}}\put(0.70537458,0.83012507){\color[rgb]{0,0,0.05882353}\makebox(0,0)[t]{\lineheight{1.25}\smash{\begin{tabular}[t]{c}(\#15) Illuminating Clay\end{tabular}}}}\put(0.38354508,0.83012504){\color[rgb]{0,0,0}\makebox(0,0)[t]{\lineheight{1.25}\smash{\begin{tabular}[t]{c}Deformable continuous tangibles \cite{ishii2012radical}\end{tabular}}}}\put(0,0){\includegraphics[width=\unitlength,page=23]{cross-classification.pdf}}\put(0.3829486,0.64871995){\color[rgb]{0,0,0}\makebox(0,0)[t]{\lineheight{1.25}\smash{\begin{tabular}[t]{c}Augmented everyday objects\end{tabular}}}}\put(0.70537458,0.67874031){\color[rgb]{0,0,0.06666667}\makebox(0,0)[t]{\lineheight{1.25}\smash{\begin{tabular}[t]{c}(\#27) Teegi\end{tabular}}}}\put(0.70484449,0.58845438){\color[rgb]{0,0,0.06666667}\makebox(0,0)[t]{\lineheight{1.25}\smash{\begin{tabular}[t]{c}(\#25) Slurp\end{tabular}}}}\put(0.70534865,0.70898544){\color[rgb]{0,0,0.06666667}\makebox(0,0)[t]{\lineheight{1.25}\smash{\begin{tabular}[t]{c}(\#12) musicBottles\end{tabular}}}}\put(0,0){\includegraphics[width=\unitlength,page=24]{cross-classification.pdf}}\put(0.70537458,0.95266166){\color[rgb]{0,0,0.05882353}\makebox(0,0)[t]{\lineheight{1.25}\smash{\begin{tabular}[t]{c}(\#26) Relief\end{tabular}}}}\put(0,0){\includegraphics[width=\unitlength,page=25]{cross-classification.pdf}}\put(0.3818888,0.75436472){\color[rgb]{0,0,0.01176471}\makebox(0,0)[t]{\lineheight{1.25}\smash{\begin{tabular}[t]{c}Mirrored manipulative artifacts\end{tabular}}}}\put(0.70537458,0.79973249){\color[rgb]{0,0,0.05882353}\makebox(0,0)[t]{\lineheight{1.25}\smash{\begin{tabular}[t]{c}(\#5) Head Prop\end{tabular}}}}\put(0.70537458,0.76903783){\color[rgb]{0,0,0.05882353}\makebox(0,0)[t]{\lineheight{1.25}\smash{\begin{tabular}[t]{c}(\#10) Voodoo Dolls\end{tabular}}}}\put(0.70537458,0.73879262){\color[rgb]{0,0,0.06666667}\makebox(0,0)[t]{\lineheight{1.25}\smash{\begin{tabular}[t]{c}(\#23) ArcheoTUI\end{tabular}}}}\put(0.70538898,0.8599207){\color[rgb]{0,0,0.03529412}\makebox(0,0)[t]{\lineheight{1.25}\smash{\begin{tabular}[t]{c}(\#28) SoundFORMS\end{tabular}}}}\put(0,0){\includegraphics[width=\unitlength,page=26]{cross-classification.pdf}}\put(0.38342311,0.79973249){\color[rgb]{0,0,0.01176471}\makebox(0,0)[t]{\lineheight{1.25}\smash{\begin{tabular}[t]{c}Mirrored everyday objects \end{tabular}}}}\put(0,0){\includegraphics[width=\unitlength,page=27]{cross-classification.pdf}}\put(0.38188881,0.54353359){\color[rgb]{0,0,0}\makebox(0,0)[t]{\lineheight{1.25}\smash{\begin{tabular}[t]{c}Manipulative artifacts\end{tabular}}}}\put(0,0){\includegraphics[width=\unitlength,page=28]{cross-classification.pdf}}\put(0.02219738,0.98265992){\color[rgb]{0,0,0}\makebox(0,0)[lt]{\lineheight{1.25}\smash{\begin{tabular}[t]{l}\cite{ishii2008beyond}\end{tabular}}}}\put(0.02219738,0.46669388){\color[rgb]{0,0,0}\makebox(0,0)[lt]{\lineheight{1.25}\smash{\begin{tabular}[t]{l}\cite{ishii2008beyond}\end{tabular}}}}\put(0.02219738,0.04468995){\color[rgb]{0,0,0}\makebox(0,0)[lt]{\lineheight{1.25}\smash{\begin{tabular}[t]{l}\cite{ishii2008beyond}\end{tabular}}}}\put(0.26977349,0.6489074){\color[rgb]{0,0,0}\makebox(0,0)[lt]{\lineheight{1.25}\smash{\begin{tabular}[t]{l}\cite{ishii2008beyond}\end{tabular}}}}\put(0.02219738,0.25769933){\color[rgb]{0,0,0}\makebox(0,0)[lt]{\lineheight{1.25}\smash{\begin{tabular}[t]{l}\cite{ishii2008beyond}\end{tabular}}}}\put(0.26977349,0.21103865){\color[rgb]{0,0,0}\makebox(0,0)[lt]{\lineheight{1.25}\smash{\begin{tabular}[t]{l}\cite{ishii2008beyond}\end{tabular}}}}\end{picture}\endgroup  }
  \caption{Matching previous classifications with what--how tangibility classes through a collection of applications.}
  \Description{The diagram is composed of four columns: (1) Genre, (2) Subgenre, (3) Application, and (4) Tangibility Class. From left to right, five genres of applications are linked to thirteen subgenres, which are linked to the 33 applications. Then, the application are linked to their corresponding classes.}
  \label{fig:cross}
\end{figure*}

Some previous classifications have discriminated between tangible user interfaces according to their morphology (\ie{} how they are physically shaped, such as surfaces \cite{ishii2008beyond}, everyday objects \cite{ishii2008beyond}, or discrete tangibles \cite{ishii2012radical}) and behavior (\ie{} how users interact with them and how they interact with users, such as assembly \cite{ishii2008beyond}, deformation \cite{ishii2012radical}, or transformation \cite{ishii2012radical}). \autoref{fig:cross} bridges these previous classifications and the tangibility classes through the collection of applicative tangible user interfaces.

Matching previous classifications, the applications' tangibility classes are split into five clusters. The present collection of applications already shows that previous classifications and tangibility classes have no direct correspondence. Whereas ``Artifacts \& Objects'' gathers applications coming from the four classes, ``Ambient Media'' and ``Constructive Assemblies'' count only one class each. Moreover, this initial matching may evolve when more applications are included. Indeed, even if some genres fit implicitly into certain classes well, based on their definition (\eg{} interactive surfaces, such as tabletops, falign well with Class~II and Class~III), some classes may appear in certain clusters. For example, constructive assemblies could also be stackable audio or video filters that are tools controlling intangible data (\ie{} describing such an application specimen would then require bridging Class~III with ``Constructive Assemblies''). Furthermore, other clusters may appear because certain genres of tangible user interfaces are absent from the collection of applications (\eg{} ``Tangible Telepresence'' \cite{ishii2008beyond} and ``Tangibles with Kinetic Memory'' \cite{ishii2008beyond}). However, we believe that the descriptive power of the four classes may provide a sufficient design space to integrate any new applicative tangible user interface and related physical user interfaces.

\section{Limitations and Future Work}\label{sec-limitations-future-work}

A limitation of this work is that analyses were not consolidated with the specimens' designers. Describing the \purple{33} applications was mostly straightforward, but sometimes required advanced questioning.
An example is when describing the Pinwheels specimen (\numpinwheels{}), which represents information flow in an ambient manner. At first glance, Pinwheels must be datnible: airflow represents the datum flow. However, this idea evolved from using airflow in the ambientROOM \cite{ishii1998ambientroom}---merely needing air coming into the room through a hole---to using visual ``spinning pinwheels'' \cite{dahley1998ambient,ishii2001pinwheels,wisneski1998pinwheels}. Hence, Pinwheels are described as datibles\footnote{Treating single datnibles (airflow) as physical user interfaces would lead to reconsidering Class~I definition or to creating a new class.}.
Another example lies in the vertical measurement graphs (\ie{} usage rate, response time, and running cost graphs) placed next to the IP Network Design Workbench (\numipworkbench{}). Some could have considered these graphs as auxiliary GUIs. However, they were considered as opnibles (\ie{} intangible measurement operations), as well as measurements in the Urp (\numurp{}) that were considered as operations.
Advanced questioning can become tricky when design thoughts are not clearly detailed in the articles. The community must benefit from future work that describes previous physical user interface specimens in a publicly shared database (\eg{} \cite{fleck2018cladistics}) and future articles that describe specimens using holistic terminologies.

This work focused on the roles found in tangible and related physical user interfaces without considering remote collaboration conditions \cite{brave1998remote}. However, in such conditions, remote users may be represented by the user interface through streams \cite{ishii1990teamworkstation,ishii1992clearboard,leithinger2014inform}, shapes \cite{brave1997intouch,brewer2007nimio}, or avatars \cite{lee2025since} and even interact with physical entities \cite{brave1997intouch,leithinger2014inform}. Therefore, these conditions raise the question of defining a role for remote users specifically (or considering them merely as data or tools?), refining their body parts (\eg{} head, forearms, and hands), and defining terms that describe how they appear in the user interface. These questions may extend the terminology to interactions with robots and cobots in Human-Robot Interaction \cite{stoeva2024body} and Extended Reality conditions \cite{gonzalez2025exploring}.

The description of bodied entities is refined on only two levels of embodiment (\ie{} tangible and graspable), which roughly simplifies the twenty values of Fishkin' taxonomy \cite{fishkin2004taxonomy}. However, this article considered a level of embodiment for bodiless entities (\ie{} intangible), even if intangible entities can also encounter various levels of representativeness (\eg{} a beep or slang audio feedback when sending a file to trash, versus paper-creasing noise).

This work separated hallmarks into classes. However, defining a distance calculation between hallmarks could serve the needs of automatic classification algorithms (\eg{} cladistics \cite{fleck2018cladistics}) that would help sort, organize, and browse the vast collection of specimens provided by the field of tangible user interfaces over the past three decades.

Finally, this article introduces the concept of interactional entities briefly, which future work will need to define more extensively.

\section{Conclusion}\label{sec-conclusion}

This article introduces a ``what--how'' terminology (WHT) that provides twelve terms to name physical representations and controls: by taking a role-centered viewpoint (\ie{} the digital role), the resulting terms self-contain meaning about how digital entities come into the physical world. Those terms are blended words, whose logical structure from known words must make them easy to learn, recollect, and understand. Mainly, this new terminology outperforms the previous terms found in the literature for three reasons. First, it defines all terms within a single namespace, free from specific paradigms, thus allowing for cross-genre descriptions. Second, it refines the descriptive level of physical user interface components. Third, it better aligns with definitions, thereby enhancing communication and understanding. This terminology introduces the term \g{datible,} which matches the definition of pure tangible user interfaces, along with providing eleven terms that distinguish other representations. This terminology must benefit at least three scopes: first, supporting ideation by providing vocabulary that describes possibilities; second, offering a holistic conceptual space for toolkit architectures; and finally, offering a taxonomy usable for the exploration and classification of physical user interfaces.

\begin{acks}
The author thanks the reviewers of the ACM CHI 2023 conference for their valuable comments, which helped revise and improve certain sections of this article, as well as extend its contributions.
\end{acks}

\bibliographystyle{ACM-Reference-Format}
\bibliography{paper.bib}

\end{document}